\documentclass[aps,prb,twocolumn,superscriptaddress]{revtex4-2}
\usepackage{graphicx}
\usepackage{bm}
\usepackage{amsmath}
\usepackage{amssymb}
\usepackage[version=4]{mhchem}
\usepackage{multirow}
\usepackage{hyperref}
\hypersetup{
colorlinks = true,
urlcolor = blue,
linkcolor = blue,
citecolor = blue
}
\begin{document}
\title{
  Theoretical proposal of superconductivity in hole-doped reduced bilayer nickelate La$_3$Ni$_2$O$_6$: a manifestation of orbital-space bilayer model with incipient bands
}
\author{Shu Kamiyama}
\email{kamiyama@presto.phys.sci.osaka-u.ac.jp}
\affiliation{Department of Physics, The University of Osaka, Toyonaka, Osaka 560-0043, Japan}
\author{Reo Kohno}
\affiliation{Faculty of Engineering, Tottori University, 4-101 Koyama-cho Minami, Tottori, Tottori 680-8552, Japan}
\author{Yuto Hoshi}
\affiliation{Faculty of Engineering, Tottori University, 4-101 Koyama-cho Minami, Tottori, Tottori 680-8552, Japan}
\author{Kensei Ushio}
\affiliation{Faculty of Engineering, Tottori University, 4-101 Koyama-cho Minami, Tottori, Tottori 680-8552, Japan}
\author{Daiki Nakaoka}
\affiliation{Faculty of Engineering, Tottori University, 4-101 Koyama-cho Minami, Tottori, Tottori 680-8552, Japan}
\author{Hirofumi Sakakibara}
\affiliation{Faculty of Engineering, Tottori University, 4-101 Koyama-cho Minami, Tottori, Tottori 680-8552, Japan}
\affiliation{Advanced Mechanical and Electronic System Research Center(AMES),
Tottori University, 4-101 Koyama-cho Minami, Tottori, Tottori 680-8552, Japan}
\author{Kazuhiko Kuroki}
\affiliation{Department of Physics, The University of Osaka, Toyonaka, Osaka 560-0043, Japan}
\date{\today}
\begin{abstract}
  A correspondence exists between the multi-orbital Hubbard model and the bilayer Hubbard model, in which superconductivity is optimized in an incipient-band regime in both cases.
  In the multi-orbital system, the orbital level offset $\Delta E$ plays a role analogous to the interlayer hopping in bilayer systems, and superconductivity is enhanced for large $\Delta E$.
  We refer to such a multi-orbital model as an orbital-space bilayer model (OSBM).
  In this study, we theoretically propose that a reduced bilayer nickelate La$_3$Ni$_2$O$_6$ can be a candidate for a superconductor described by OSBM when an appropriate amount of holes is doped.
  By constructing a tight-binding model based on first-principles calculations, a large $\Delta E$ between the Ni $d_{x^2-y^2}$ and the other $d$ orbitals is obtained due to the absence of outer apical oxygens.
  Furthermore, our fluctuation exchange approximation calculations indicate the emergence of $s\pm$-wave superconductivity driven by interorbital interactions in an incipient-band situation, where the superconducting gap function changes its sign between the $d_{x^2-y^2}$ and other $d$ orbital bands.
  We also investigate the energetic and dynamical stability of the crystal structure under atomic substitution and pressure.
  Although La$_3$Ni$_2$O$_7$ and La$_3$Ni$_2$O$_6$ share a similar chemical formula, our study shows that an entirely different pairing mechanism can take place in the latter.
\end{abstract}
\maketitle

\section{introduction}
\begin{figure*}
  \centering
  \includegraphics[width=0.95\linewidth]{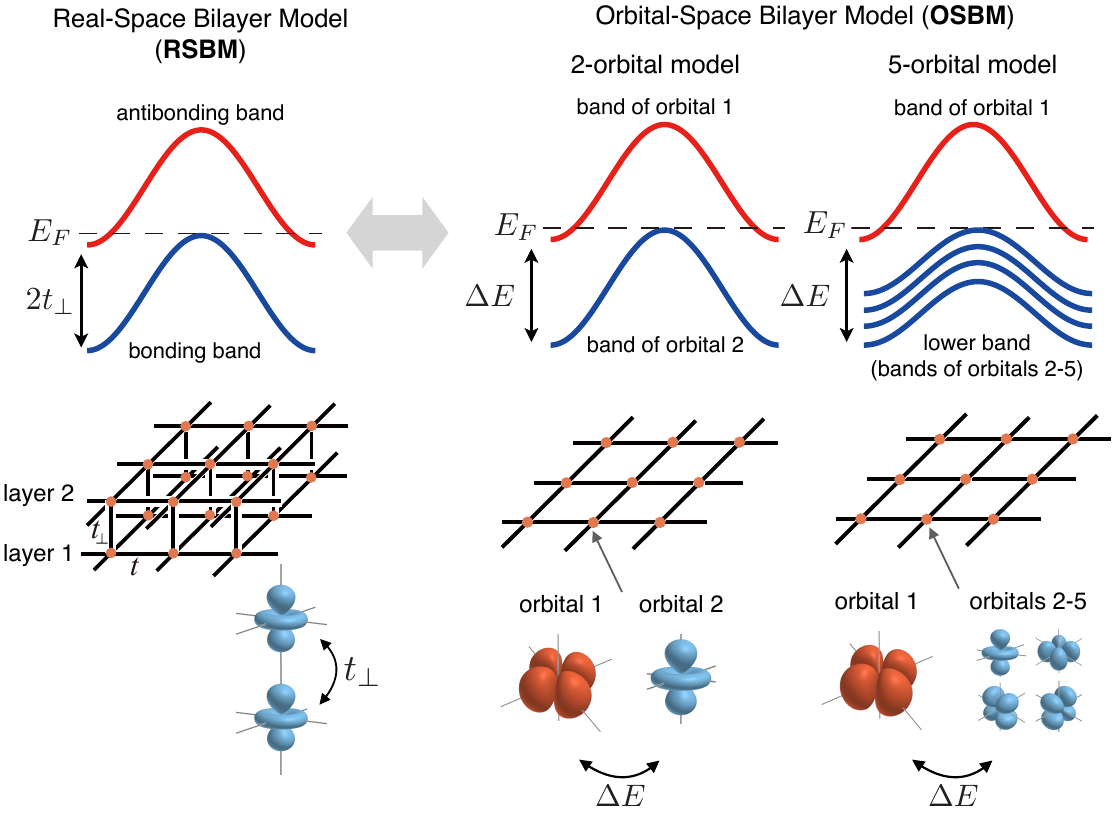}
  \caption{
    Schematics of the Real-Space Bilayer Model (RSBM) and the Orbital-Space Bilayer Model (OSBM).
    The interlayer hopping $t_\perp$ in the RSBM gives rise to bonding and antibonding bands with an energy separation of $2t_\perp$.
    In the OSBM, orbital-level offsets $\Delta E$ between multiple orbitals play an analogous role.
    In both models, superconductivity is optimized in the incipient-band regime, where one band intersects the Fermi level while the other touches or lies slightly away from it.
    The correspondence between the RSBM and the OSBM has been established not only in a two-orbital model but also in multi-orbital systems, such as the five-orbital model.
    }
  \label{fig_intro}
\end{figure*}
Nickelate superconductors have recently attracted considerable attention due to the discovery of two different types.
First, in the infinite-layer nickelates \ce{$Ln$NiO2}~\cite{D_Li_2019,Y_Nomura_2022,S_Chow_2025}, superconductivity with a transition temperature of $T_c \sim$ 10--40~K has been achieved. 
The \ce{Ni^{1+}} state realizes a $d^{9-\delta}$ electronic configuration, for which several theories have suggested that $d$ wave-like superconductivity arises similarly to the cuprates~\cite{H_Sakakibara_2020,X_Wu_2020,M_Kitatani_2020,M_Kitatani_2023}.
Second, in the bilayer Ruddlesden-Popper-type nickelate \ce{La3Ni2O7}, superconductivity with $T_c \sim 80$~K has been realized under pressures of 15--40~GPa~\cite{H_Sun_2023,M_Wang_2024}, attracting considerable attention as a new class of high-$T_c$ superconductors.
Recently, a variation of this material \ce{(La,Sm)3Ni2O7} has attained a $T_c$ of 96~K~\cite{F_Li_2026}.
Also, ambient-pressure superconductivity has been found in thin films of \ce{(La,Pr)3Ni2O7} with a $T_c$ of 40~K~\cite{E_Ko_2025,G_Zhou_2025} to 60~K~\cite{G_Zhou_2025_2}. 

In these bilayer nickelates, the inner apical oxygen mediates strong interlayer hopping $t_{\perp}$ between the $d_{3z^2-r^2}$ orbitals~\cite{M_Nakata_2017}.
This is in contrast to the cuprates, where superconductivity arises primarily from a single \ce{CuO2} plane and is well described by single-orbital $d$-wave pairing, with orbital hybridization acting to suppress superconductivity~\cite{H_Sakakibara_2010,H_Sakakibara_2012}.
However, it has been theoretically pointed out that there is an upper limit to the transition temperature $T_c$ in single-layer and single-orbital systems such as cuprates~\cite{P_Monthoux_1999,R_Arita_2000,R_Arita_2001}.
Therefore, the bilayer Hubbard model has been theoretically explored as a platform for realizing high-$T_c$ superconductivity with transition temperatures exceeding those of the cuprates~\cite{bilayer0,bilayer1,bilayer2,bilayer3,bilayer4,bilayer5,bilayer6,bilayer7,bilayer8,bilayer9,bilayer10,bilayer11}.
Several studies, including the work by one of the present authors, have demonstrated that a bilayer system with strong interlayer hopping $t_\perp$ and near half-filling can give rise to a superconducting transition temperature beyond that of conventional cuprates through the formation of interlayer pairing~\cite{bilayer6, bilayer11,Y_Nomura_2025}.
In fact, prior to the experimental discovery of superconductivity, \ce{La3Ni2O7} was considered by one of the present authors as a possible candidate material that can realize such a situation~\cite{M_Nakata_2017}.

In a bilayer model, interlayer hopping $t_\perp$ gives rise to bonding and antibonding bands originating from the same orbitals, resulting in an energy splitting of $2t_\perp$, and superconductivity is enhanced for large $t_\perp$ and near half-filling (see the left of Fig.~\ref{fig_intro})~\cite{bilayer6, bilayer11,Y_Nomura_2025}.
In this regime, either the bonding or antibonding band touches or nearly touches the Fermi level, while the other lies slightly below or above it, thereby realizing an incipient-band situation.
In multiband systems, it has been shown that superconductivity is optimized in the presence of an incipient band~\cite{P_Hirschfeld_2011,Y_Bang_2016,Y_Bang_2019,K_Kuroki_2005, D_Ogura_2017, K_Matsumoto_2018, K_Matsumoto_2020, H_Sakamoto_2020, D_Kato_2020, T_Aida_2024, T_Yagi_2024}.
(Here we use the term ``incipient band'' in a wide sense. Namely, we include situations where the Fermi level intersects the band at energies close to the band edge.)
The bilayer model is also one of such systems in which superconductivity is enhanced by an incipient band~\cite{M_Nakata_2017}.

On the other hand, our previous studies have shown that the multi-orbital Hubbard model is another example in which superconductivity is enhanced in the presence of an incipient band~\cite{K_Yamazaki_2020,N_Kitamine_2020,N_Kitamine_2023,H_Sakakibara_2025}.
Although the bilayer Hubbard model involves only a single orbital with intraorbital interactions, the two-orbital Hubbard model comprises two orbitals per site and incorporates interorbital interactions.
Despite these differences, in both models, superconductivity is optimized in the incipient-band regime~\cite{K_Yamazaki_2020}.
The two models are mathematically equivalent when $U=U'=J=J'$ as shown in Ref.~\cite{H_Shinaoka_2015}, where onsite interactions $U$, $U'$, $J$, and $J'$ are the intraorbital repulsion, interorbital repulsion, Hund's coupling, and pair hopping, respectively, with $2t_\perp$ in the bilayer model corresponding to the orbital level offset $\Delta E$ in the two-orbital model~(see the middle of Fig.~\ref{fig_intro}).
In reality, $U > U'> J \sim J'$ and interorbital hybridization is present; nevertheless, our calculations in the two-orbital Hubbard model for a new type of cuprate \ce{Ba2CuO_{3+\delta}} in Ref.~\cite{K_Yamazaki_2020} indicate that the correspondence between the bilayer and two-orbital models persists to some extent, and that superconductivity is enhanced with increasing $\Delta E$.
Furthermore, we have found that the correspondence between the two models persists not only in the two-orbital model but also in the five-orbital model~\cite{N_Kitamine_2020,N_Kitamine_2023,H_Sakakibara_2025}. 
Our calculations have shown that, even in the five-orbital system, superconductivity is enhanced by increasing the level offset $\Delta E$ between one orbital and the remaining four orbitals, and is optimized in the incipient-band regime~\cite{N_Kitamine_2020,N_Kitamine_2023,H_Sakakibara_2025}~(see the right of Fig.~\ref{fig_intro}).

Based on the correspondence between the bilayer and multi-orbital Hubbard models, we hereafter refer to the former as the real-space bilayer model (\textbf{RSBM}) and the latter as the orbital-space bilayer model (\textbf{OSBM}).
$2t_\perp$ in the RSBM corresponds to $\Delta E$ in the OSBM.

Although superconductors based on the OSBM have not yet been reported experimentally, we have theoretically suggested some candidate materials, one of which is heavily hole-doped infinite-layer nickelate \ce{(La,Sr)NiO2} with a $d^{8+\delta}$ configuration~\cite{N_Kitamine_2020, H_Sakakibara_2025}.
As mentioned above, $d$-wave superconductivity in infinite-layer nickelates with a $d^{9-\delta}$ electronic configuration has already been suggested theoretically, analogous to conventional cuprates~\cite{H_Sakakibara_2020,X_Wu_2020,M_Kitatani_2020,M_Kitatani_2023}.
On the other hand, we have focused on the large orbital-level offset $\Delta E$ between $d_{x^2-y^2}$ and other Ni-$d$ orbitals, arising from the \ce{NiO2} square planar structure.
In Ref.~\cite{H_Sakakibara_2025}, the possibility of superconductivity in a state close to the $d^{8+\delta}$ electron configuration has been evaluated, which might be realized by heavy hole doping of the $d^9$ parent compound.
As a result, we have theoretically predicted the possibility of $s\pm$-wave superconductivity mediated by interorbital interaction in the OSBM mechanism.
We have also predicted that $T_c$ could potentially exceed that of the conventional infinite-layer nickelates with a $d^{9-\delta}$ electronic configuration.

Along this line of consideration, the reduced bilayer nickelate \ce{La3Ni2O6}~\cite{V_Poltavets_2006,V_Poltavets_2009,N_Warren_2013,Z_Liu_2022,A_Botana_2016,P_Worm_2022,Y_Zhang_2024} could also become one of the candidates for OSBM superconductors.
In a series of $n$-layer reduced Ruddlesden-Popper nickelates \ce{La$_{n+1}$Ni$_n$O$_{2(n+1)}$}, $n=2$ corresponds to \ce{La3Ni2O6}, and $n=\infty$ corresponds to the infinite-layer compound \ce{LaNiO2}.
Since both compounds have a square planar structure, the large $\Delta E$ in \ce{LaNiO2}~\cite{H_Sakakibara_2025} is also expected to be present in \ce{La3Ni2O6}.
As the number of layers $n$ decreases, the $d$ orbital occupation is reduced, thereby approaching a $d^8$ electronic configuration.
This trend is consistent with our proposal in Ref.~\cite{H_Sakakibara_2025}.
Since \ce{La3Ni2O6} has a $d^{8.5}$ configuration, the realization of a $d^{8+\delta}$ state via hole doping is easier than in infinite-layer nickelates.

\begin{figure}[h]
  \centering
  \includegraphics[width=0.95\linewidth]{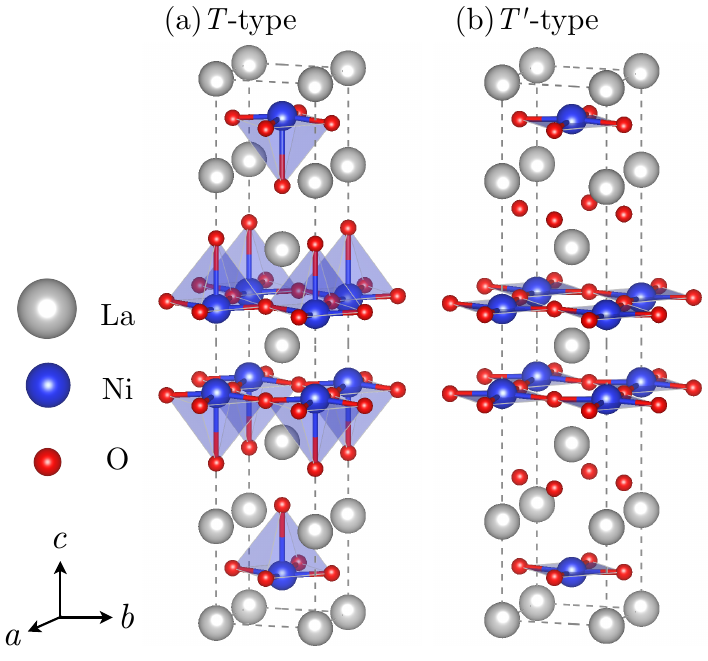}
  \caption{The crystal structure of \ce{La3Ni2O6}. (a) and (b) indicate the $T$- and $T'$-type structures, respectively.}
  \label{fig:cryst}
\end{figure}

\ce{La3Ni2O6} has already been reported to be synthesized by reducing \ce{La3Ni2O7} with \ce{CaH2}~\cite{V_Poltavets_2006}.
While the calculated band structure is metallic within the generalized gradient approximation (GGA)~\cite{V_Poltavets_2009}, \ce{La3Ni2O6} exhibits nontrivial insulating behavior, with an energy gap of approximately 100~meV experimentally confirmed at 300~K~\cite{Z_Liu_2022}.
Although the electrical resistivity shows metallic behavior at pressures above 6.1~GPa, superconductivity has not yet been observed experimentally~\cite{Z_Liu_2022}.
The crystal structure adopts the $T'$-type structure~(see Fig.~\ref{fig:cryst}(b)); however, structural transitions, such as a transition to the $T$-type structure, may also occur through atomic substitution or by applying pressure as in cuprates~\cite{H_Muller_1977}.

Theoretically, the possibility of superconductivity in \ce{La3Ni2O6} has been analyzed in previous studies.
In Ref.~\cite{Y_Zhang_2024}, it was concluded that $s\pm$ pairing (based on the RSBM in our terminology) as in \ce{La3Ni2O7} is unlikely due to the weak interlayer coupling of the $d_{3z^2-r^2}$ orbitals.
On the other hand, in Ref.~\cite{F_Lechermann_2024}, the possibility of RSBM-based $s\pm$ superconductivity was proposed using the sic-DMFT band structure.

In this paper, we focus on hole-doped \ce{La3Ni2O6} as a new candidate for an OSBM superconductor.
We construct tight-binding models based on first-principles calculations and analyze superconductivity within the fluctuation-exchange (FLEX) approximation.
This paper is organized as follows.
Methods for first-principles calculations and the FLEX approximation are described in Sec.~\ref{sec:method}. 
In Secs.~\ref{subsec:str_param} and \ref{subsec:band}, we optimize the crystal structures of \ce{La3Ni2O6} for both the $T$- and $T'$-type structures and construct five-orbital models based on first-principles calculations, employing three different treatments of electronic correlation effects: GGA, GGA+$U$, and Quasi-particle Self-consistent GW (QSGW).
In Sec.~\ref{subsec:SC}, we investigate superconductivity using a combination of FLEX and the linearized Eliashberg equation, and show that superconductivity can be enhanced in the large $\Delta E$ and incipient-band regime.
In Sec.~\ref{subsec:int_compare}, we analyze the role of interorbital interactions within the OSBM framework by comparing two different models.
In Sec.~\ref{subsec:stability}, we also evaluate the stability of the crystal structures under pressure and atomic substitution by calculating phonon dispersions or comparing total enthalpies.
Discussions are given in Sec.~\ref{sec:discussion}.
Sec.~\ref{sec:conclusion} summarizes the study.

\section{method\label{sec:method}}
First-principles calculations based on the density functional theory (DFT) are performed by using the PBEsol~\cite{PBEsol_1,PBEsol_2} exchange-correlation functional as implemented in Vienna {\it ab initio} simulation package (VASP)~\cite{vasp1,vasp2,vasp3,vasp,vasp4}. 
We use a plane-wave cutoff energy of 600 eV for Kohn-Sham orbitals and an $12\times 12\times 12$ Monkhorst-Pack ${\bm k}$ mesh.
We perform structural optimization under the PBEsol functional until the Hellmann-Feynman force becomes less than 0.01 eV \AA$^{-1}$ for each atom. 
Both the lattice parameters and atomic coordinates are optimized for the tetragonal structure (space group: $I4/mmm$). 
Based on the optimized structure, we perform the first-principles band calculation using PBEsol and PBEsol$+U$ functional, where we set $U=3$~eV for Ni-3$d$ orbitals.
In addition to the GGA calculations, we use QSGW~\cite{Kotani2014PMT-QSGW, Kotani2010PMT, Kotani2007QSGW} method using the ecalj package~\cite{ecalj} to check the validity of the results of GGA$+U$ method. 
In this paper, we employ a hybrid method between QSGW and GGA, named QSGW80~\cite{Deguchi2016QSGW80}, for exchange correlation functional in order to correct possible overestimations of the band gap size. 
We take an $8\times 8 \times 8$ $\bm{k}$ mesh, sufficiently large for QSGW.

From the calculated band structures, we extract the Wannier functions~\cite{wannier_1,wannier_2} using the WANNIER90~\cite{wannier90} codes and construct a two-site five-orbital model consisting of all the Ni $3d$ orbitals.
We also perform constrained random phase approximation (cRPA) calculations to determine the Coulomb and exchange interaction parameters, using an $8\times 8 \times 8$ $\bm{k}$ mesh within the GGA.

We analyze superconductivity based on the obtained five-orbital Hubbard models within the FLEX approximation~\cite{FLEX_1,FLEX_2}.
We calculate the self-energy induced by the spin-fluctuation formulated as shown in the literature and obtain the renormalized Green's function by solving the Dyson's equation in a self-consistent calculation.
The obtained Green's function and the pairing interaction, mediated mainly by spin fluctuations, are plugged into the linearized Eliashberg equation.
Since the eigenvalue $\lambda$ of the equation reaches unity at $T= T_c$, here we adopt $\lambda$, obtained at a fixed temperature of $T$ = 0.01 eV, to measure how close the system is to superconductivity.
For convenience, we will call the eigenfunction (with the largest eigenvalue) of the linearized Eliashberg equation at the lowest Matsubara frequency ($i\pi k_B T $) the superconducting gap function.
We take a $16 \times 16 \times 4$ $\bm{k}$ mesh and 2048 Matsubara frequencies for the FLEX calculation.

To investigate the dynamical stability of the crystal structure, we perform phonon calculation using the frozen-phonon method as implemented in Phonopy package~\cite{phonopy_1,phonopy_2}
We take a $2\times 2\times 2$ $\bm{q}$ mesh with a $6\times 6\times 6$ $\bm{k}$ mesh unless noted and investigate whether imaginary modes appear or not.

\section{Results}
\begin{figure*}[htbp]
  \centering
  \includegraphics[width=0.8\linewidth]{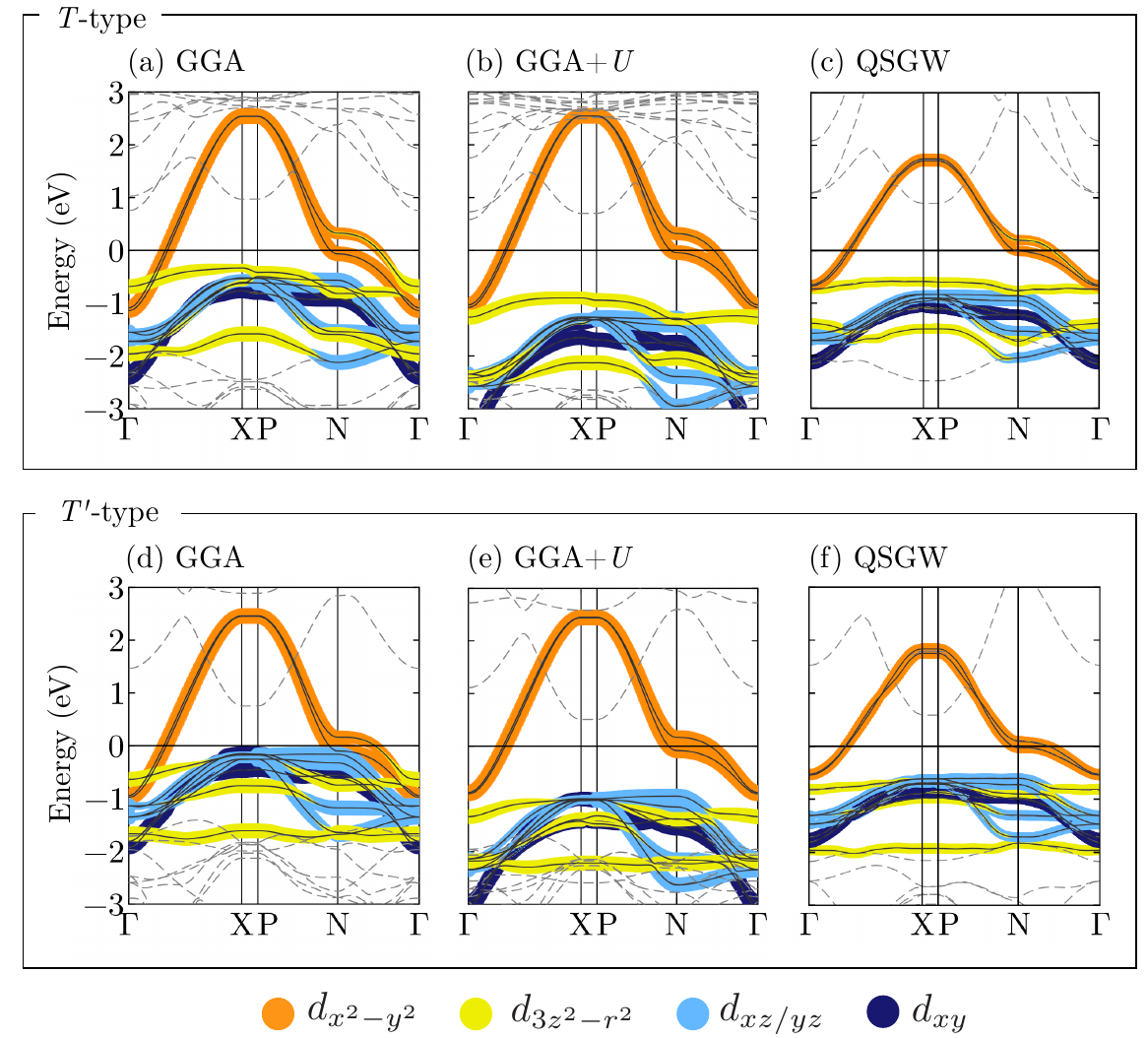}
  \caption{
    The band structures of \ce{La3Ni2O6} in (a)--(c) the $T$-type and (d)--(f) the $T'$-type structure.
    The band dispersion of the five-orbital model is superposed to the first-principles bands.
  }
  \label{fig_band}
\end{figure*}

\subsection{Crystal structural parameters \label{subsec:str_param}}
We perform structural optimization of \ce{La3Ni2O6} for both the $T$ and $T'$ structures.
The obtained lattice constants and volumes are $a=3.760$~\AA, $c=19.88$~\AA~($V=281.9$~\AA$^3$) for the $T$ structure and $a=3.898$~\AA, $c=19.33$~\AA~($V=293.7$~\AA$^3$) for the $T'$ structure.
The interlayer Ni--Ni distance is 3.09~\AA\ and 3.04~\AA\ in the $T$ and $T'$ structures, respectively.
In \ce{La3Ni2O7}, it is approximately 4~\AA~\cite{H_Sun_2023}, while in \ce{La3Ni2O6}, the absence of the inner apical oxygen results in a shorter Ni--Ni distance.
The Ni--O--Ni angle in the \ce{NiO2} plane is 169.6$^\circ$ for the $T$ structure and 173.9$^\circ$ for the $T'$ structure, indicating that the $T$ structure exhibits larger buckling than the $T'$ structure.
The Ni--O distance for the outer apical oxygen in the $T$ structure is 2.56~\AA, which is larger than the value of approximately 2~\AA\ reported for \ce{La3Ni2O7}~\cite{H_Sun_2023}.

\subsection{Electronic band structure \label{subsec:band}}
\begin{table}[h]
  \centering
  \begin{tabular}{ccccc}
  \hline\hline
  $T$ (GGA$+U$) & $d_{x^2-y^2}$ & $d_{3z^2-r^2}$ & $d_{xz/yz}$ & $d_{xy}$ \\ 
  $\Delta E$ & -----         & 2.18           & 2.51           & 2.56     \\ 
  $t$      & $-$0.43         & $-$0.01          & $-$0.31          & $-$0.19    \\ 
  $t_{\perp}$          & $-$0.04         & $-$0.49          & 0.12          & $-$0.05    \\ \hline
  $T'$ (GGA$+U$) & $d_{x^2-y^2}$ & $d_{3z^2-r^2}$ & $d_{xz/yz}$ & $d_{xy}$ \\ 
  $\Delta E$ & -----         & 2.36           & 2.25           & 2.28     \\ 
  $t$      & $-$0.40         & 0.01          & $-$0.32          & $-$0.17    \\ 
  $t_{\perp}$          & $-$0.04         & $-$0.45          & 0.13          & $-$0.04    \\ \hline\hline
  \end{tabular}
  \caption{
    Parameters of the tight-binding model obtained from the constructed five-orbital model for the $T$ and $T'$ structures based on the GGA$+U$ band dispersions.
    $\Delta E$ indicates the onsite energy difference between the $d_{x^2-y^2}$ orbital and the other $d$ orbitals.
    $t$ and $t_{\perp}$ indicate the intralayer nearest-neighbor and interlayer hopping, respectively.
    The units of all parameters are eV.
  }
  \label{tablH_TBparams}
\end{table}

Using the optimized crystal structure, band calculations are performed for each of the $T$ and $T'$ structures using three methods: GGA, GGA$+U$ $(U=3\,\textrm{eV})$, and QSGW.
The GGA$+U$ and QSGW calculations are performed to evaluate the correlation effects.
The QSGW method provides a high-precision independent quasiparticle picture without empirical parameters, based on the GW approximation derived from many-body perturbation theory.
An orbital level offset $\Delta E$ is crucial in OSBM. 
To assess the validity of the value of $U$ in GGA$+U$, we perform band calculations using the QSGW method.
We then construct a five-orbital tight-binding model with all Ni $3d$ orbitals from the obtained band structures.

Figure~\ref{fig_band} shows the band structure obtained from first-principles calculations and the five-orbital model.
A large orbital level offset $\Delta E$ is observed between the $d_{x^2-y^2}$ orbital and the other $d$ orbitals across the bands.
Compared to the GGA bands [Figs.~\ref{fig_band}(a) and \ref{fig_band}(d)], $\Delta E$ is enhanced in the GGA$+U$ bands [Figs.~\ref{fig_band}(b) and \ref{fig_band}(e)].
The band structures calculated using QSGW [Figs.~\ref{fig_band}(c) and \ref{fig_band}(f)] agree well with those obtained within GGA$+U$ [Figs.~\ref{fig_band}(b) and \ref{fig_band}(e)], indicating that the choice of $U=3\,\mathrm{eV}$ is reasonable.
Note that the bandwidth is reduced in QSGW by correcting the self-interaction error in GGA due to the nonlocal nature of the dynamically screened exchange interaction~\cite{Q_Jang_2015,H_Yuto_2026}.
In the following sections, we will proceed with the analysis on the basis of the GGA and GGA$+U$ band structures.

Parameters of the tight-binding models derived from the GGA$+U$ bands are summarized in Table~\ref{tablH_TBparams}~(Parameters from the GGA and QSGW bands are shown in Appendix~\ref{appendix:parameter}).
In the $T'$ structure, the level offsets $\Delta E$ between the $d_{x^2-y^2}$ orbital and the other $d$ orbitals are nearly the same across the $d_{3z^2-r^2}$, $d_{xz/yz}$, and $d_{xy}$ orbitals.
On the other hand, in the $T$ structure, $\Delta E_{3z^2-r^2}$, the level offset between the $d_{3z^2-r^2}$ and $d_{x^2-y^2}$ orbitals, is smaller than the corresponding offsets for the other $d$ orbitals.
This reduction compared to the $T'$ structure originates from the presence of the outer apical oxygen in the $T$ structure, which raises the energy level of the $d_{3z^2-r^2}$ orbital.
However, the value of $\Delta E_{3z^2-r^2}$ in the $T$-type \ce{La3Ni2O6} is still larger than that in \ce{La3Ni2O7}~\cite{H_Sakakibara_2024}, which can be attributed to the absence of the inner apical oxygen and the larger Ni-O distance for the outer apical oxygen in \ce{La3Ni2O6}.
The interlayer hopping $t_\perp$ of the $d_{3z^2-r^2}$ orbital exhibits smaller values compared to that in \ce{La3Ni2O7}~\cite{H_Sakakibara_2024}, again due to the absence of the inner apical oxygen, but still possesses appreciable values, which would be due to the closer interlayer Ni--Ni distance.

In the band structure shown in Fig.~\ref{fig_band}, a splitting appears in the $d_{x^2-y^2}$ band near the Fermi level at the N point.
This is a bilayer splitting originating from the large $t_{\perp}$ of the $d_{3z^2-r^2}$ orbital and the hybridization between the $d_{x^2-y^2}$ and $d_{3z^2-r^2}$ orbitals.
The possibility of superconductivity due to RSBM arising from this bilayer splitting, which has some relation to the previous theoretical studies in Ref.~\cite{Y_Zhang_2024, F_Lechermann_2024}, will be discussed in Sec.~\ref{sec:discussion}.
It should be noted that, within the present band structure, the $d_{3z^2-r^2}$ orbital alone does not satisfy the conditions for RSBM superconductivity~\cite{bilayer6}, because the orbital is nearly fully occupied.

\subsection{Superconductivity \label{subsec:SC}}
\begin{figure*}[tbp]
 \centering
 \includegraphics[width=1.0\linewidth]{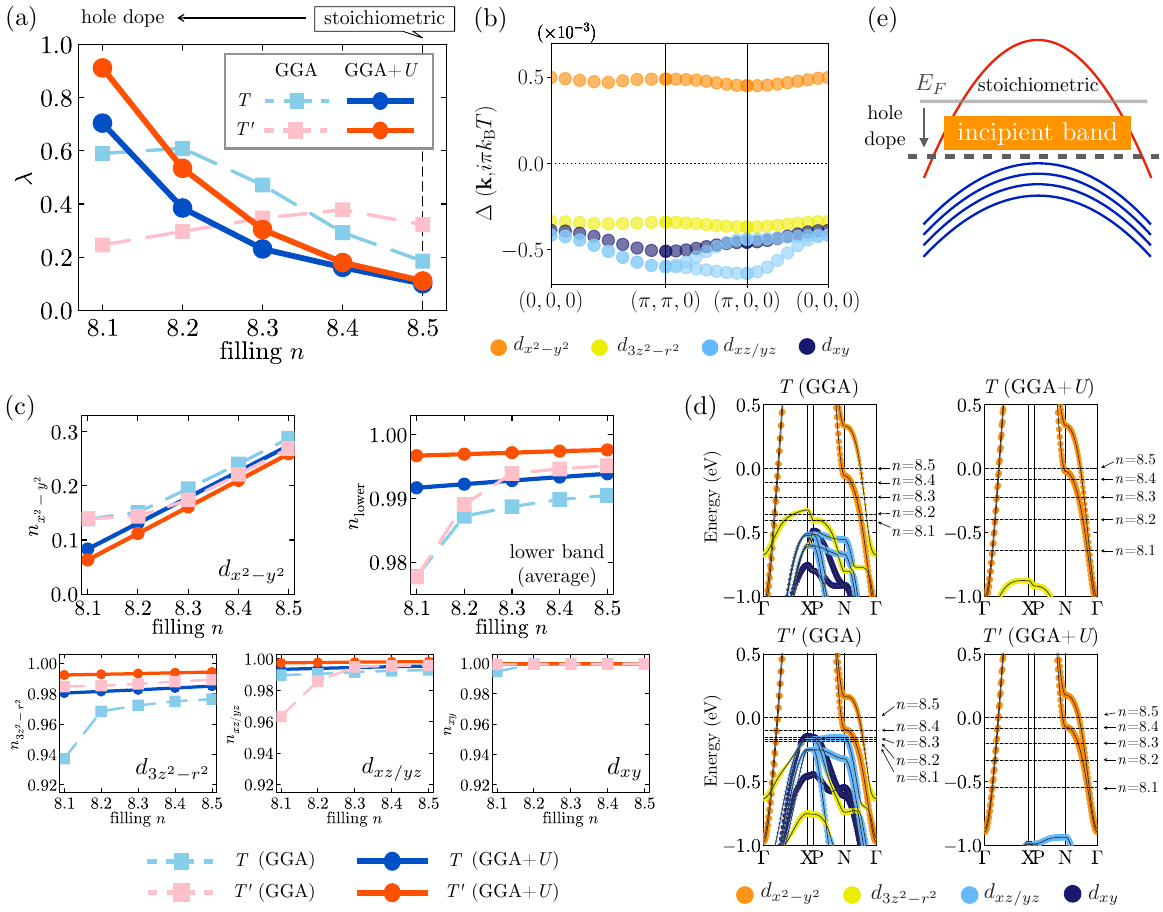}
 \caption{
  (a) The eigenvalues $\lambda$ of the linearized Eliashberg equation as a function of the band filling $n$ (the number of electrons per Ni atom), where $n=8.5$ corresponds to the stoichiometric case.
  The blue (red) lines are for the $T$ ($T'$) structure, and the dashed (solid) lines with squares (circles) represent the values obtained from the GGA (GGA$+U$) band structures, respectively.
  (b) Gap function obtained from the GGA$+U$ bands of the $T'$ structure ($n=8.1$).
  (c) Orbital occupations as functions of the Ni $d$ orbital filling. The ``lower band'' represents the average occupation of the $d_{3z^2-r^2}$, $d_{xz/yz}$, and $d_{xy}$ orbitals.
  (d) The filling dependence of the chemical potential $\mu$.
  Black dashed lines indicate the level of the chemical potential $\mu$ for each band filling from 8.1 to 8.5. We set the value of $\mu$ at $n=8.5$ to zero.
  (e) Schematics of the realization of incipient bands by hole doping.
}
 \label{fig_SC}
\end{figure*}

\begin{figure*}[tbp]
 \centering
 \includegraphics[width=1.0\linewidth]{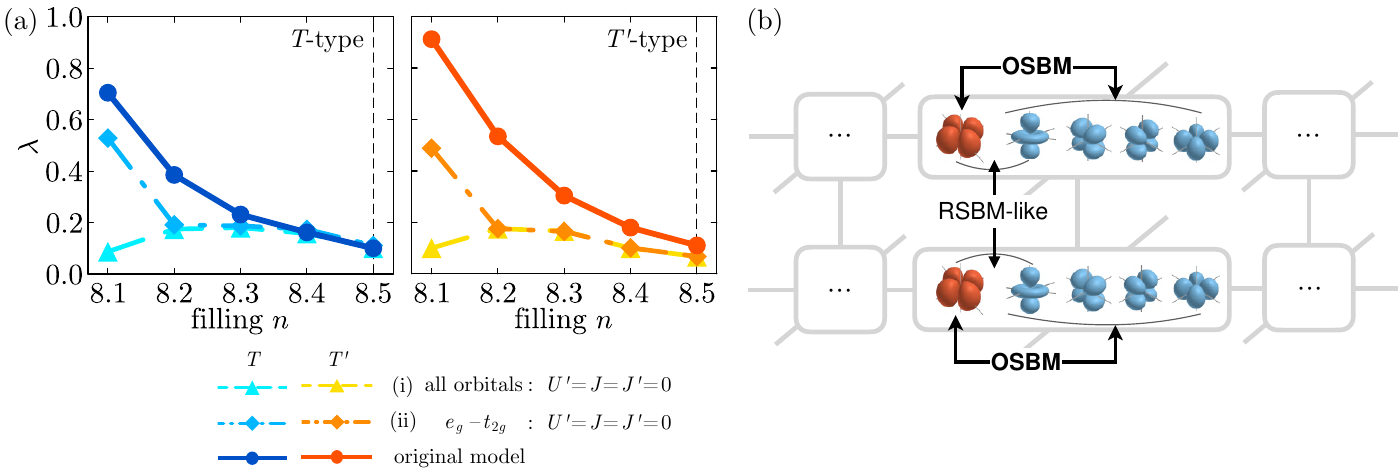}
 \caption{
  (a) Filling dependence of the eigenvalue $\lambda$ of the Eliashberg equation when the interorbital interactions.
  In addition to the calculations with the interaction parameters obtained from cRPA (denoted as ``original''), results are also shown for (i) $U'=J=J'=0$ for all orbitals and (ii) $U'=J=J'=0$ between $e_g$ and $t_{2g}$ orbitals.
  (b) Relationship between the orbitals at each Ni site in \ce{La3Ni2O6} and the OSBM and RSBM.
  Within the present model and analysis, the OSBM contribution dominates over that of the RSBM (see the text for details).
 }
 \label{fig_interorbital}
\end{figure*}

We now move on to the FLEX calculation for superconductivity.
We analyze the linearized Eliashberg equation for both the $T$ and $T'$ structures as a function of band filling within the rigid-band picture.
The interaction parameters are obtained from cRPA calculations, as shown in Appendix~\ref{appendix:parameter}.

The eigenvalue $\lambda$ of the linearized Eliashberg equation against the band filling for $8.1\leq n \leq 8.5$ is shown in Fig.~\ref{fig_SC}(a), where $n=8.5$ corresponds to the stoichiometric case.
Fig.~\ref{fig_SC}(b) shows the gap function obtained using the GGA$+U$ band in the case of $n=8.1$ for the $T'$ structure.
It demonstrates that the lower four bands other than the $d_{x^2-y^2}$ band have a gap with the same sign, which is opposite to that of the $d_{x^2-y^2}$ band.
In this sense, the pairing corresponds to an $s\,\pm$-wave state, indicating superconductivity arising in OSBM.
Throughout the cases studied in the present work, the gap function that has the largest eigenvalue exhibits this $s\pm$ structure.

Focusing on Fig.~\ref{fig_SC}(a), for the case of the GGA band, the eigenvalue $\lambda$ exhibits a peak of $\lambda=0.61$ at $n=8.2$ for the $T$ structure and $\lambda = 0.38$ at $n=8.4$ for the $T'$ structure.
In the GGA$+U$ case, $\lambda$ takes values of $0.70$ and $0.91$ at $n=8.1$ for the $T$ and $T'$ structures, respectively, which are the largest within the range of our calculations.
Since $n=8.0$ corresponds to the $d^8$ electronic configuration, which tends to form either a high-spin antiferromagnetic insulator or a low-spin band insulator, we exclude this case from our calculations.
These results indicate that hole doping promotes superconductivity in \ce{La3Ni2O6}.
Note that the value of $\lambda$ and the position of the peak involve a certain degree of uncertainty, since the choice of the value of $U$ in the GGA$+U$ calculation is somewhat arbitrary.

To clarify why hole doping enhances superconductivity in \ce{La3Ni2O6}, we present the filling dependence of the orbital-resolved occupations in Fig.~\ref{fig_SC}(c) and that of the chemical potential in the tight-binding model in Fig.~\ref{fig_SC}(d).
First, we focus on the occupation of each orbital in Fig.~\ref{fig_SC}(c).
In the GGA case, for fillings below $n=8.3$, the occupations of the four orbitals $d_{3z^2-r^2},d_{xz/yz}, d_{xy}$ (hereafter referred to as the lower bands) decrease, indicating the occurrence of hole doping~(see Fig.~\ref{fig_SC}(c)).
On the other hand, in the GGA$+U$ case, while the lower bands remain nearly fully filled, the occupation of the $d_{x^2-y^2}$ orbital approaches zero. 
Next, let us focus on the chemical potential in Fig.~\ref{fig_SC}(d).
We find that, at fillings where $\lambda$ reaches its peak, the top of the lower band touches or lies slightly below the Fermi level, such that the lower band is fully filled while the $d_{x^2-y^2}$ band (upper band) is nearly empty, i.e., an incipient-band situation is realized.
As mentioned in the Introduction, superconductivity in the OSBM is optimized in the incipient-band regime.
While the stoichiometric \ce{La3Ni2O6} in the $d^{8.5}$ configuration is not suitable for superconductivity, hole doping drives the system into an incipient-band regime, thereby enhancing superconductivity~(see Fig.~\ref{fig_SC}(e)).

\subsection{Comparison of the effects of the interorbital interaction \label{subsec:int_compare}}
Next, we investigate the effects of interorbital interactions on the OSBM.
In this section, we focus exclusively on the GGA$+U$ bands.
In the former section, we use a model including interorbital interactions with $U, U', J, J'$ obtained via cRPA method.
In addition to the original model, we consider two modified models to investigate the effects of interorbital interactions on superconductivity in the OSBM.
In these models, interorbital interactions are selectively eliminated as follows:
(i) $U'=J=J'=0$ for all orbitals;
(ii) $U'=J=J'=0$ only between the $e_g$ and $t_{2g}$ orbitals.
The former is intended to examine how superconductivity changes when the effects of the OSBM are entirely removed.
In the latter model, since the $d_{x^2-y^2}$ orbital interacts only with the $d_{3z^2-r^2}$ orbital, we focus on isolating the effects of the OSBM arising from the interorbital interaction between the $d_{x^2-y^2}$ and $d_{3z^2-r^2}$ orbitals.

Fig.~\ref{fig_interorbital}(a) shows the eigenvalue of the linearized Eliashberg equation for the above three choices of interaction parameters.
$\lambda$ takes small values at $n=8.5$ in any case, which corresponds to the stoichiometric situation.
Model (i) exhibits a moderate peak around $n=8.3$ in both the $T$ and $T'$ structures.
This can be attributed to an RSBM-like pairing effect arising from the bilayer splitting of the $d_{x^2-y^2}$ band at the N point, as mentioned in Sec.~\ref{subsec:band} (see Fig.~\ref{fig_band}).
Here, we refer to the situation as ``RSBM-like'' because the main orbital component is the $d_{x^2-y^2}$ orbital, with a little hybridization of $d_{3z^2-r^2}$, as illustrated in Fig.~\ref{fig_interorbital}(b).
In any case, within our model and analysis, the RSBM-like interlayer pairing instability is found to be weak, which is consistent with a previous theoretical study~\cite{Y_Zhang_2024}.
On the other hand, Ref.~\cite{F_Lechermann_2024} proposed an RSBM scenario in which the $d_{3z^2-r^2}$ orbital plays a central role in this material.
Since their analysis is based on sic-DMFT and RPA methods, a direct comparison with our results obtained using GGA$+U$ and FLEX is not straightforward.

In the range $8.2\leq n \leq 8.4$, models (i) and (ii) show good agreement, whereas the original model, in which all interorbital interactions are taken into account, exhibits larger values of $\lambda$.
This indicates that, in this range, the interorbital interaction between the $d_{x^2-y^2}$ and $d_{3z^2-r^2}$ orbitals alone is insufficient to enhance the OSBM effect.
Instead, the OSBM behavior emerges from interactions between the $t_{2g}$ and $e_g$ orbitals, namely, interorbital interactions between the $d_{x^2-y^2}$ orbital and the other four orbitals ($d_{3z^2-r^2}$, $d_{xy}$, $d_{xz}$, and $d_{yz}$), as shown in Fig.~\ref{fig_interorbital}(b).

For $n = 8.1$, the $\lambda$ value decreases in model (i) but increases in model (ii), differing from that of the original model.
This behavior implies that, in this regime, the RSBM effect becomes weakened as the Fermi level drops below the bilayer splitting, while the OSBM effect instead emerges through the interorbital interaction between the $d_{x^2-y^2}$ and $d_{3z^2-r^2}$ orbitals.
The significant increase in $\lambda$ in the original model toward $n = 8.1$, where all interorbital interactions are taken into account, originates from the Fermi level approaching the $d_{3z^2-r^2}$ and the $t_{2g}$ bands, rendering these bands incipient, as discussed in Sec.~\ref{subsec:SC} [Fig.~\ref{fig_SC}(c)].

\begin{figure}[t]
  \centering
  \includegraphics[width=1.0\linewidth]{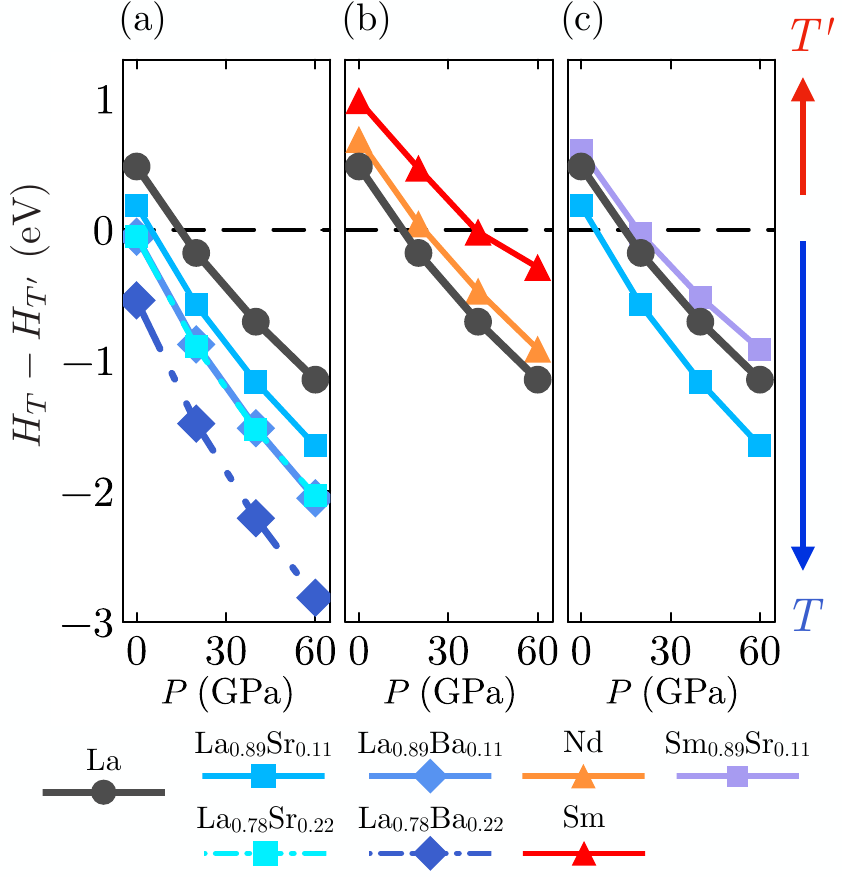}
  \caption{
  Enthalpy difference between the $T$ and $T'$ structures of \ce{$A$3Ni2O6} under various pressures and $A$-site substitutions.
  (a) Alkaline-earth-metal doping, $A = \mathrm{La}_{1-x}{AE}_x$ (${AE}=\mathrm{Sr}, \mathrm{Ba}$; $x = 0.11, 0.22$).
  (b) Lanthanide substitution, $A = \mathrm{Nd}, \mathrm{Sm}$.
  (c) Lanthanide-substituted alkaline-earth doping, $A = \ce{Sm_{0.89}Sr_{0.11}}$ (with $A = \ce{La_{0.89}Sr_{0.11}}$ for comparison).
  }
  \label{fig_thermal}
\end{figure}

\subsection{Stability of the crystal structure \label{subsec:stability}}
In this section, we investigate the stability of the crystal structure of \ce{La3Ni2O6} under atomic substitution and hydrostatic pressure.
\ce{La3Ni2O6} adopts the $T'$ structure at ambient pressure~\cite{V_Poltavets_2006}.
However, it is well known that cuprates can transform into different structures, such as $T$ and $T'$, depending on the choice of the $A$-site element~\cite{H_Muller_1977}.
Therefore, it is reasonable to expect that \ce{La3Ni2O6} may also undergo a structural transition induced by $A$-site substitution.
Furthermore, it has been reported that the electrical conductivity changes from insulating to metallic under pressure~\cite{Z_Liu_2022}.
Motivated by the bilayer nickelate \ce{La3Ni2O7}, which exhibits superconductivity accompanied by a structural transition under pressure~\cite{H_Sun_2023}, we also explore the possibility of pressure-induced structural transitions.

\subsubsection{Energetic stability: comparison of the enthalpy between the $T$ and $T'$ structures\label{subsec_enthalpy}}
To evaluate the energetic stability between the $T$ and $T'$ structures, we calculate the total enthalpy difference between the two structures under pressure and atomic substitution.
We define the enthalpy difference between the two structures as $\Delta H = H_T - H_{T'}$. 
A positive value of $\Delta H$ ($\Delta H > 0$) indicates that $H_T > H_{T'}$, and thus the $T'$ structure is energetically more stable, whereas a negative value ($\Delta H < 0$) implies that $H_T < H_{T'}$ and the $T$ structure is more stable.

We begin by examining the effect of hydrostatic pressure on the parent compound \ce{La3Ni2O6}.
While the $T'$ structure has a lower energy at ambient pressure, consistent with experimental observations~\cite{V_Poltavets_2006}, the $T$ structure becomes energetically more stable as pressure increases.
Next, we consider the hole-doped case realized by substituting Sr or Ba at the $A$ site, i.e., \ce{(La_{1-$x$}$AE$_{$x$})_3Ni2O6} ($AE=$~Sr, Ba), with substitution levels of $x = 0.11$ and $0.22$, as shown in Fig.~\ref{fig_thermal}(a).
In all the cases shown in Fig.~\ref{fig_thermal}(a), hole doping shifts the enthalpy curves toward the regime in which the $T$ structure becomes more stable.
Moreover, Ba substitution stabilizes the $T$ structure more effectively than Sr substitution, and this stabilization is further enhanced at higher hole-doping levels.
In particular, at $P=0$~GPa, the $T$ structure is more stable in $A=$ \ce{La_{0.78}Sr_{0.22}}, \ce{La_{0.89}Ba_{0.11}}, and \ce{La_{0.78}Ba_{0.22}}.
This suggests that hole doping can induce a structural transition from the $T'$ to the $T$ structure even at ambient pressure.

On the other hand, we also consider the case where the La atom at the $A$ site is replaced by other lanthanide elements.
Fig.~\ref{fig_thermal}(b) shows the enthalpy difference for $A=$ La, Nd, and Sm.
In contrast to the hole-doped case, lanthanide substitution shifts the enthalpy curves toward stabilizing the $T'$ structure, with the magnitude of the shift increasing in the order La$ \to$ Nd $\to$ Sm.

These results can be understood in terms of an internal pressure effect arising from differences in ionic radii.
In Fig.~\ref{fig_thermal}(a), the ionic radii of \ce{La^{3+}}, \ce{Sr^{2+}}, and \ce{Ba^{2+}} are 1.17, 1.32, and 1.49 \AA~\cite{R_Shannon_1976}, respectively.
The shift toward the $T$ structure becomes stronger as the ionic radius increases.
Furthermore, as shown in Fig.~\ref{fig_thermal}(b), the ionic radius decreases along the lanthanide series from La to Nd to Sm, and correspondingly, the stability shifts toward the $T'$ structure.
In all cases, external pressure shifts the enthalpy curve toward the region where the $T$ structure is more stable.
This suggests that physical external pressure or internal pressure arising from ionic size stabilizes the $T$ structure, whereas reducing these pressure-like effects stabilizes the $T'$ structure.

Additionally, we calculate the enthalpy difference for $A=$ \ce{Sm_{0.89}Sr_{0.11}} to examine the effect of hole doping in the lanthanide-substituted system (see the purple line in Fig.~\ref{fig_thermal}(c)).
Compared with the case of $A=\ce{La_{0.89}Sr_{0.11}}$, the enthalpy curve shifts toward stabilizing the $T'$ structure.
This is thought to be the effect of reduced internal pressure due to the smaller ionic radius of \ce{Sm^{3+}} compared to \ce{La^{3+}}.
These results suggest that lanthanide substitution may lead to energetic stabilization of the $T'$ structure even under hole-doping conditions.

Note that, to investigate the effect of chemical pressure on changes in the lattice constants, we present the lattice constants under pressure and atomic substitution in Appendix~\ref{appendix:lat}.
In the Sr-doped cases, the lattice constants are shorter than in the non-doped case, whereas in the Ba-doped cases the lattice constants become longer.
However, the magnitude of the lattice constants does not necessarily correlate with the stability of the $T$ and $T'$ structures.
We therefore consider that the internal pressure effect arising from differences in ionic radii has a greater impact on the energy changes than the chemical pressure associated with variations in the lattice constants.

\begin{figure*}[tb]
  \centering
  \includegraphics[width=1.0\linewidth]{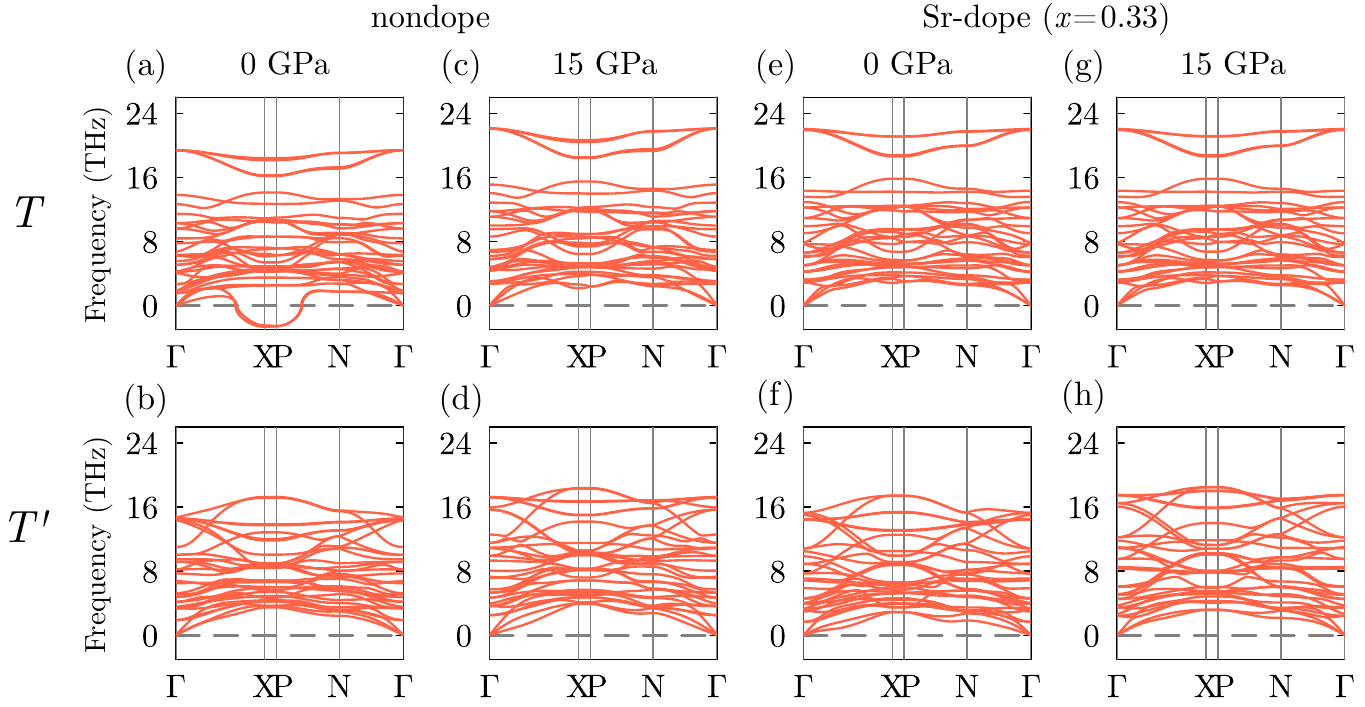}
  \caption{
    The phonon dispersions of (a)--(d) non-doped \ce{La3Ni2O6} and (e)--(h) Sr-doped \ce{(La_{1-x}Sr_x)_3Ni2O6} ($x=0.33$) for both the $T$ and $T'$ structures under pressures of $P=0$ and 15~GPa.  
  }
  \label{fig_phonon}
\end{figure*}

\subsubsection{Dynamical stability: phonon dispersion \label{subsec_phonon}}
Next, to examine the dynamical stability of the crystal structure, we calculated the phonon dispersions and checked for the presence of imaginary modes.
We calculated the phonon dispersions of non-doped \ce{La3Ni2O6} and Sr-doped \ce{(La_{1-$x$}Sr_x)3Ni2O6} ($x=0.33$) for both the $T$ and $T'$ structures under pressures of $P=0$ and 15~GPa, showing the results in Fig.~\ref{fig_phonon}.
Note that only for the $T'$ structure at $P=0$~GPa, shown in Fig.~\ref{fig_phonon}(b), we use a $3\times 3\times 3$ $\bm{q}$ mesh with a $4\times 4\times 4$ $\bm{k}$ mesh for accuracy.

First, we consider the non-doped state, shown in the left panel of Fig.~\ref{fig_phonon}.
Focusing on the results at ambient pressure~[Figs.~\ref{fig_phonon}(a) and \ref{fig_phonon}(b)], no imaginary modes appear in the $T'$ structure, whereas they are present in the $T$ structure. 
This is consistent with experimental observations that the $T'$ structure is synthesized at ambient pressure, confirming that the $T'$ structure is indeed stable~\cite{V_Poltavets_2006}.
In contrast, under a pressure of 15~GPa (where the $T$ structure can be energetically more stable), no imaginary modes appear in either the $T$ or the $T'$ structure, indicating that both are dynamically stable.
Note that the $T$ structure is energetically stable under 15~GPa, although dynamical stability is also indicated in the $T'$ structure. 

Next, we turn to the Sr-doped state, shown in the right panel of Fig.~\ref{fig_phonon}.
Although the $T$ structure is energetically more stable in this region based on the result from Sec.~\ref{subsec_enthalpy}, none of the results, including the $T'$ structure cases, exhibit imaginary modes, indicating dynamical stability.
These results indicate that, even in the region where the $T$ structure is energetically more stable, the $T'$ structure can be dynamically stable. 
This suggests that when the enthalpy difference between the two structures is small, the $T'$ structure might be obtained as a metastable phase, even though the $T$ structure is energetically favored.

\section{Discussion\label{sec:discussion}}
Here, let us compare the band calculation results for \ce{La3Ni2O6} obtained using different methods.
In the spectral function calculated by DMFT~\cite{P_Worm_2022}, the $t_{2g}$ orbitals shift upward and form the Fermi surface.
In contrast, in the spectrum obtained using sic-DMFT, as described in Ref.~\cite{F_Lechermann_2024}, the $t_{2g}$ orbitals shift downward away from the Fermi level, while the $d_{3z^2-r^2}$ orbital shifts upward.
In our GGA$+U$ calculations, as well as in Ref.~\cite{Y_Zhang_2024}, the $t_{2g}$ and $d_{3z^2-r^2}$ orbitals shift to lower energies, moving away from the Fermi level.
A similar tendency is also observed in the QSGW band structure obtained in this paper.
In addition, we calculate the spectral function obtained from the GGA$+U+$FLEX method, as shown in Appendix~\ref{appendix:spectral}.
These results indicate that the behavior of the bands varies significantly depending on the method used.
At the moment, it is difficult to determine which method provides the correct direction of the correlation-driven band shift, and experimental observations such as ARPES may be necessary.

Despite its metallic DFT band structure, \ce{La3Ni2O6} is insulating at ambient pressure and exhibits an anomaly in the electrical resistivity around 100~K.
According to Ref.~\cite{Z_Liu_2022}, an energy gap of 54~meV was observed in the temperature range $84 \leq T \leq 125\,\mathrm{K}$, while a larger gap of 100~meV was observed for $176 \leq T \leq 300\,\mathrm{K}$.
In the sic-DMFT results of Ref.~\cite{F_Lechermann_2024}, a gap of approximately 50~meV is successfully reproduced in calculations at 50~K.
Although our present approach cannot deal with this insulating state for the stoichiometric case, we believe that once the system is metallized by doping, the present results become applicable.

It is known that metallization occurs under pressures of 6.1~GPa or higher in \ce{La3Ni2O6}, but superconductivity has not been observed under pressures up to 25.5~GPa~\cite{Z_Liu_2022}.
In fact, this is consistent with our calculations showing that superconductivity is unlikely to emerge in the stoichiometric state ($n=8.5$), even if the system is made metallic.
We have shown in the present study that hole doping induces an incipient-band state, in which superconductivity is expected to be enhanced.
However, there remains an unfavorable possibility that the insulating state persists and superconductivity does not emerge even upon hole doping.
In this case, applying hydrostatic pressure might provide an effective route to suppress the insulating state and realize a superconducting state.
Here, adopting small-radius $A$-site elements such as Sm may be favorable for maintaining the $T'$ structure.
On the other hand, the hole-doped $T$ structure under pressure itself may result in superconductivity, as can be seen from a relatively large $\lambda$ even for the $T$ structure (see Fig.~\ref{fig_SC}(a)), which is a consequence of the large $\Delta E$ due to the absence of the inner apical oxygen and large Ni--O distance for the outer apical oxygen.

\section{Conclusion\label{sec:conclusion}}
In this paper, we have investigated the possibility of superconductivity based on OSBM in the reduced bilayer nickelate \ce{La3Ni2O6}.
We consider two types of structures, $T$ and $T'$, and performed band-structure calculations using three methods: GGA, GGA$+U$, and QSGW.
We have shown that a large orbital level offset $\Delta E$ exists between the $d_{x^2-y^2}$ orbital and the other Ni-$d$ orbitals.

By analyzing superconductivity within FLEX calculations, we have found that $s\,\pm$-wave superconductivity can be enhanced by hole doping due to interorbital interactions in both the $T$ and $T'$ structures.
This enhancement originates from the OSBM mechanism, where the Fermi level approaches the lower bands and the incipient-band situation is realized.
Furthermore, we demonstrate that superconductivity is not significantly enhanced by the interorbital interaction between the $d_{x^2-y^2}$ and $d_{3z^2-r^2}$ orbitals alone; rather, it is enhanced by the interorbital interactions involving all four orbitals in the lower bands.

We have also shown that the energy difference between the $T$ and $T'$ structures varies under external hydrostatic pressure or internal pressure effects arising from differences in ionic radii, resulting in a shift of the stable structure.
In addition, stable phonon dispersions are obtained under pressure in both the $T$ and $T'$ structures.
Considering that the insulating behavior at ambient pressure disappears under pressure, we consider the possibility of suppressing the insulating property and realizing OSBM-like superconductivity by applying pressure and hole doping.

In OSBM, high-$T_c$ could be expected through large $\Delta E$, which would offer significant potential in material design.
Our study aims to explore superconductivity in multi-orbital systems based on the OSBM concept.
If superconductivity driven by OSBM can indeed be realized, as proposed in this work, it would provide a promising route toward enhancing transition temperatures in unconventional superconductors and expanding the design flexibility of future superconducting materials.

\begin{acknowledgments}
We thank Masataka Kakoi, Ryota Mizuno, Masanori Nagao, Masayuki Ochi, Hiroya Sakurai for fruitful discussions.
The computing resource is supported by the supercomputer system (system-B) in the Institute for Solid State Physics, the University of Tokyo.
This work was supported by Grants-in-Aid for Scientific Research from JSPS, KAKENHI Grants No.~JP24K01333, JP25H01252, JP25K00959, JP25K08459.
S.K. was supported by JST FOREST Program, Grant No. JPMJFR212P. 
H.S was supported by JST FOREST Program, Grant No. JPMJFR246T. 

\end{acknowledgments}

\appendix

\section{model parameters \label{appendix:parameter}}
In Table~\ref{tablH_TBparams} of the main text, we present the parameters of the tight-binding model constructed from the GGA$+U$ band dispersions.
Here, Table~\ref{tablH_TBparams_appendix} shows the corresponding parameters obtained from the GGA and QSGW band dispersions.
The values of $\Delta E$ in the GGA are smaller than those in GGA$+U$.
As mentioned in Sec.~\ref{subsec:band}, the bandwidths in QSGW tend to be smaller than those in GGA, and accordingly the hopping parameters take smaller values.

In this paper, we obtained the interaction parameters of onsite Coulomb repulsion $U,U'$, Hund's coupling $J$, and pair-hopping $J'$ evaluated with cRPA, as shown in Table~\ref{table_crpa}.
These are obtained based on the GGA bands since the $+U$ correction does not significantly affect the screening channels outside the target subspace.
For simplicity, we abbreviate the four-orbital-index representation of the partially screened interaction integral $V$, namely, $V_{llmm}\to U_{lm}'$ and $V_{lmlm}(=V_{lmml})\to J_{lm}$.
Note that $U_{lm}'=J_{ml}'$, $J_{lm}=J_{ml}$.

  \begin{table}[htb]
    \begin{tabular}{ccccc}
    \hline\hline
    $T$ (GGA)  & $d_{x^2-y^2}$  & $d_{3z^2-r^2}$ & $d_{xz/yz}$  & $d_{xy}$\\
    $\Delta E$ &  --  & 1.66  & 1.73  & 1.69  \\
    $t$        & $-$0.44 & $-$0.02 & $-$0.28 & $-$0.18 \\
    $t_\perp$  & $-$0.03 & $-$0.55 & 0.14  & $-$0.06 \\\hline
    $T'$ (GGA) & $d_{x^2-y^2}$  & $d_{3z^2-r^2}$ & $d_{xz/yz}$  & $d_{xy}$\\
    $\Delta E$ &  -- & 1.74  & 1.38  & 1.37\\
    $t$        & $-$0.41 & 0.01  & $-$0.27 & $-$0.16  \\
    $t_\perp$  & $-$0.04 & $-$0.50 & 0.14  & $-$0.05  \\\hline\hline

    $T$ (QSGW) & $d_{x^2-y^2}$  & $d_{3z^2-r^2}$ & $d_{xz/yz}$  & $d_{xy}$\\ 
    $\Delta E$ &  --  & 1.48  & 1.69  & 1.75  \\
    $t$        & $-$0.29 & $-$0.01 & $-$0.20 & $-$0.12 \\
    $t_\perp$  & $-$0.02 & $-$0.43 & 0.11  & $-$0.04 \\\hline
    $T'$ (QSGW) & $d_{x^2-y^2}$  & $d_{3z^2-r^2}$ & $d_{xz/yz}$  & $d_{xy}$\\
    $\Delta E$ &  -- & 1.79  & 1.47  & 1.52\\
    $t$        & $-$0.27 & 0.01  & $-$0.20 & $-$0.11  \\
    $t_\perp$  & $-$0.01 & $-$0.51 & 0.11  & $-$0.04  \\\hline\hline
    \end{tabular}
    \caption{Parameters of the tight-binding model obtained from the GGA and QSGW band dispersions.
    The units of all parameters are eV.    
    }
    \label{tablH_TBparams_appendix}
    \end{table}
    \begin{table}[ht]
      \begin{tabular}{ccccc}
      \hline\hline
      (eV)  & \multicolumn{2}{c}{$T$-type} & \multicolumn{2}{c}{$T'$-type} \\
      $l,m$ & $U,U'$        & $J,J'$       & $U,U'$        & $J,J'$        \\ \hline
      1,1   & 3.43          & --           & 3.26          & --            \\
      1,2   & 1.88          & 0.58         & 2.00          & 0.66          \\
      1,3   & 2.29          & 0.60         & 2.18          & 0.59          \\
      1,4   & 2.29          & 0.60         & 2.18          & 0.59          \\
      1,5   & 2.79          & 0.33         & 2.60          & 0.32          \\
      2,2   & 3.00          & --           & 3.62          & --            \\
      2,3   & 2.40          & 0.42         & 2.60          & 0.47          \\
      2,4   & 2.40          & 0.42         & 2.60          & 0.47          \\
      2,5   & 1.91          & 0.57         & 2.01          & 0.63          \\
      3,3   & 3.64          & --           & 3.52          & --            \\
      3,4   & 2.31          & 0.63         & 2.22          & 0.61          \\
      3,5   & 2.33          & 0.62         & 2.19          & 0.59          \\
      4,4   & 3.64          & --           & 3.52          & --            \\
      4,5   & 2.33          & 0.62         & 2.19          & 0.59          \\
      5,5   & 3.51          & --           & 3.25          & --            \\ \hline\hline
      \end{tabular}
      \caption{
        Interaction parameters of the onsite Coulomb repulsions $U$ and $U'$, Hund's coupling $J$, and pair-hopping $J'$ evaluated with cRPA.
        The orbital indices $l,m=1,2,3,4,5$ indicate the $d_{x^2-y^2}$, $d_{3z^2-r^2}$, $d_{xz}$, $d_{yz}$, and $d_{xy}$ orbitals, respectively.
        The units of all parameters are eV.
      }
      \label{table_crpa}
      \end{table}

\section{Lattice constants\label{appendix:lat}}
\begin{figure}[ht]
\centering
\includegraphics[width=0.94\linewidth]{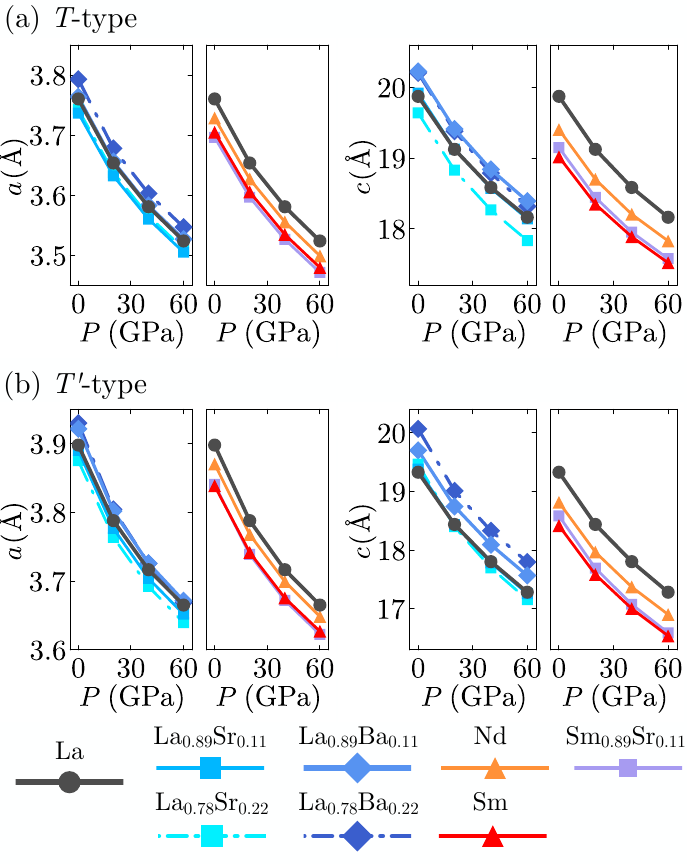}
\caption{
  The lattice constants under various pressures and for various choices of $A$-site substitution.
  Each color and marker corresponds to those used in Fig.~\ref{fig_thermal}.
}
\label{fig_lat_consts}
\end{figure}

In Fig.~\ref{fig_lat_consts}, we show the lattice constants under various pressures and for various choices of $A$-site substitution.
For Sr substitution, both the $a$ and $c$ axis lattice constants decrease slightly.
In contrast, for Ba substitution, the lattice constants increase.
For lanthanide substitutions, the lattice constants shrink in all cases.
We should note that the chemical pressure arising from changes in lattice constants and the energetic stability between the $T$ and $T'$ structures do not necessarily correspond, as mentioned in Sec.~\ref{subsec_enthalpy}.

\section{spectral function \label{appendix:spectral}}
\begin{figure}[ht]
  \centering
  \includegraphics[width=1.0\linewidth]{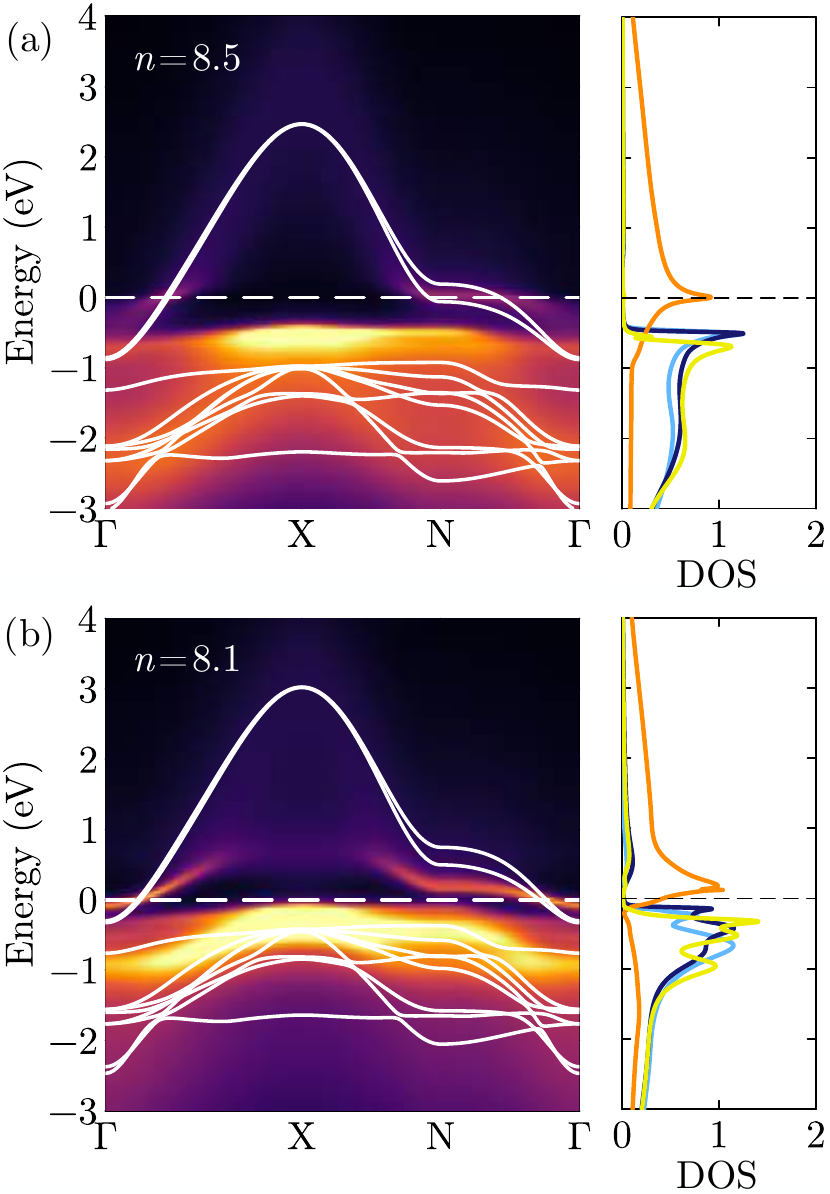}
  \caption{
    The spectral function (left) and pDOS (right) of the $T'$ structure of \ce{La3Ni2O6} obtained by GGA+$U$+FLEX calculation at (a) $n=8.5$ and (b) $n=8.1$, respectively.
    The white curves in the spectral function indicate the bare band dispersion of the tight-binding model.
    Each colored line in the pDOS corresponds to the orbital components shown in Fig.~\ref{fig_band}.
    The unit of the DOS is 1/eV per unit cell per spin.
}
  \label{fig_spectral}
\end{figure}

In Fig.~\ref{fig_spectral}, we show the GGA+$U$+FLEX spectral functions and projected density of states (pDOS) of the $T'$ structure for $n=8.5$ and $n=8.1$.
The spectral function $A(\bm{k},\omega)$ and DOS $\rho_a(\omega)$ are defined as
\begin{align*}
A(\bm{k},\omega) &= -\frac{1}{\pi} \mathrm{Im}, G(\bm{k},\omega),\\
\rho_{a}(\omega) &= \frac{1}{N_{\bm{k}}} \sum_{\bm{k}} A_{aa}(\bm{k},\omega),
\end{align*}
where $a$ denotes the orbital component, and $G(\bm{k},\omega)$ is obtained from the analytical continuation of the FLEX Matsubara Green's function using the Pad\'{e} approximation.

In the GGA+$U$+FLEX spectra, the lower bands are not shifted as far away from the Fermi level as in the GGA+$U$ band structure for both $n=8.5$ and $n=8.1$.
For $n=8.5$, the lower band lies about 0.7~eV below $E_F$.
Similarly, for $n=8.1$, the lower bands shift upward relative to those of the tight-binding model and become very close to the Fermi level, where the incipient-band situation is realized.

\section{Band dispersion of \ce{(La_{1-$x$} Sr_$x$)3Ni2O6} and weight of $p$ orbital \label{appendix:vca}}
\begin{figure*}[t]
  \centering
  \includegraphics[width=1.0\linewidth]{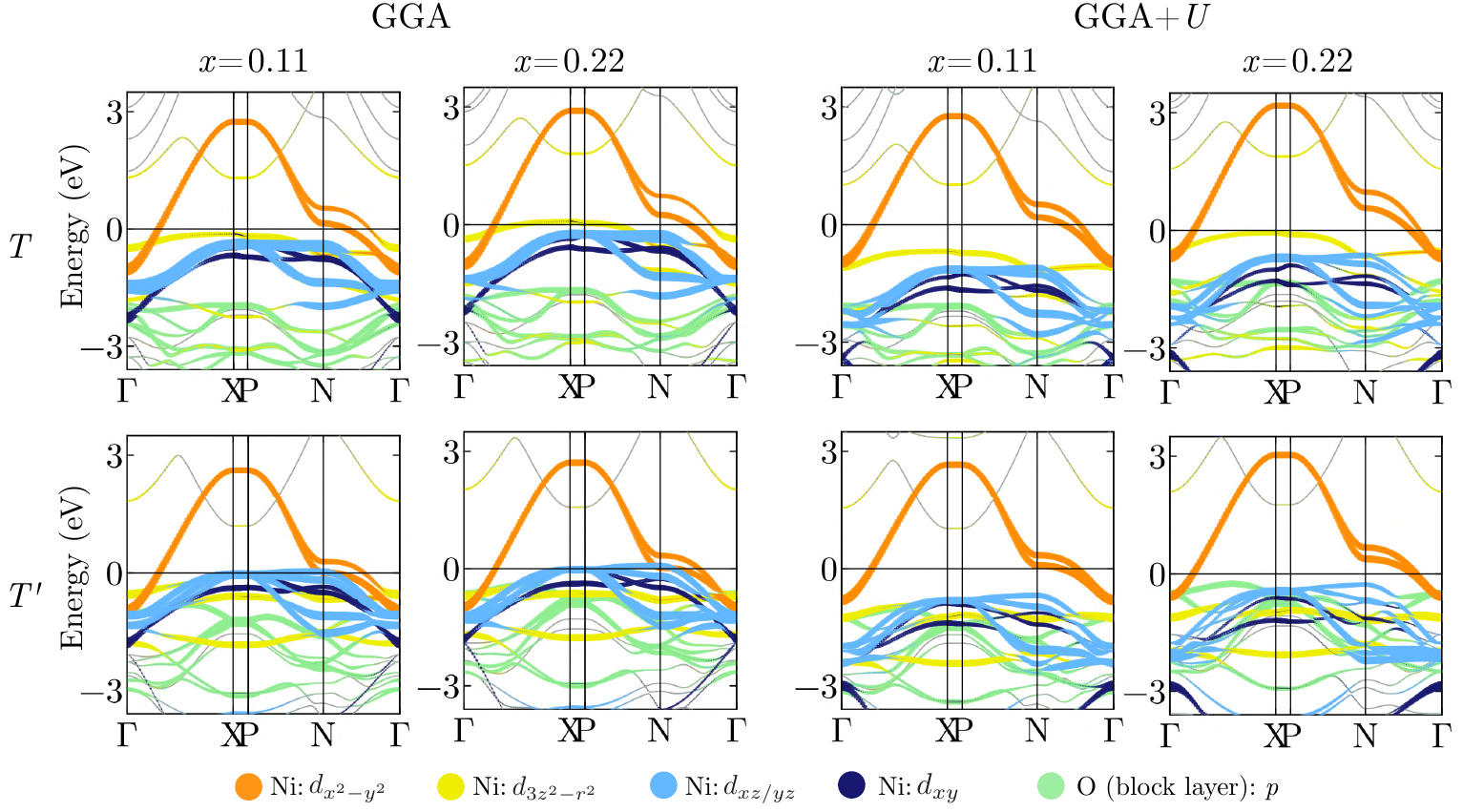}
  \caption{
    Band dispersions of \ce{(La_{1-$x$} Sr_$x$)3Ni2O6} ($x=0.11, 0.22$) for the $T$ and $T'$ structures calculated using GGA and GGA$+U$ with VCA.
    Colored lines indicate the orbital weights obtained from first-principles calculations.
    Orange, yellow, sky blue, and deep blue correspond to the Ni-$d_{x^2-y^2}$, $d_{3z^2-r^2}$, $d_{xz/yz}$, and $d_{xy}$ orbitals, respectively.
    Green lines represent the $p$ orbitals of the oxygens within the rocksalt and fluorite block layers in the $T$ and $T'$ structures, respectively.
  }
  \label{fig_vca_band}
\end{figure*}

In Fig.~\ref{fig_vca_band}, we show the band dispersions of \ce{(La_{1-$x$} Sr_$x$)3Ni2O6} ($x=$~0.11, 0.22) for the $T$ and $T'$ structures calculated using GGA and GGA$+U$ within the virtual crystal approximation (VCA).
In addition to the orbital weights of the Ni-$d$ orbitals, we also show the orbital weights of the O-$p$ orbitals in the rocksalt layer and the fluorite block layer for the $T$ and $T'$ structures, respectively, in Fig.~\ref{fig_vca_band}.
As the doping ratio $x$ increases, the Fermi level approaches or crosses the $d_{3z^2-r^2}$ or $d_{xz/yz}$ bands.
We note that in the $T'$ structure, particularly in the GGA$+U$ bands, the O-$p$ bands move closer to the Fermi level with increasing $x$.
This behavior may reflect the instability of the $T'$ structure induced by internal pressure effects associated with differences in atomic radii, as discussed in Sec.~\ref{subsec:stability}.
\bibliography{326paper}

\begin{thebibliography}{80}%
\makeatletter
\providecommand \@ifxundefined [1]{%
 \@ifx{#1\undefined}
}%
\providecommand \@ifnum [1]{%
 \ifnum #1\expandafter \@firstoftwo
 \else \expandafter \@secondoftwo
 \fi
}%
\providecommand \@ifx [1]{%
 \ifx #1\expandafter \@firstoftwo
 \else \expandafter \@secondoftwo
 \fi
}%
\providecommand \natexlab [1]{#1}%
\providecommand \enquote  [1]{``#1''}%
\providecommand \bibnamefont  [1]{#1}%
\providecommand \bibfnamefont [1]{#1}%
\providecommand \citenamefont [1]{#1}%
\providecommand \href@noop [0]{\@secondoftwo}%
\providecommand \href [0]{\begingroup \@sanitize@url \@href}%
\providecommand \@href[1]{\@@startlink{#1}\@@href}%
\providecommand \@@href[1]{\endgroup#1\@@endlink}%
\providecommand \@sanitize@url [0]{\catcode `\\12\catcode `\$12\catcode
  `\&12\catcode `\#12\catcode `\^12\catcode `\_12\catcode `\%12\relax}%
\providecommand \@@startlink[1]{}%
\providecommand \@@endlink[0]{}%
\providecommand \url  [0]{\begingroup\@sanitize@url \@url }%
\providecommand \@url [1]{\endgroup\@href {#1}{\urlprefix }}%
\providecommand \urlprefix  [0]{URL }%
\providecommand \Eprint [0]{\href }%
\providecommand \doibase [0]{https://doi.org/}%
\providecommand \selectlanguage [0]{\@gobble}%
\providecommand \bibinfo  [0]{\@secondoftwo}%
\providecommand \bibfield  [0]{\@secondoftwo}%
\providecommand \translation [1]{[#1]}%
\providecommand \BibitemOpen [0]{}%
\providecommand \bibitemStop [0]{}%
\providecommand \bibitemNoStop [0]{.\EOS\space}%
\providecommand \EOS [0]{\spacefactor3000\relax}%
\providecommand \BibitemShut  [1]{\csname bibitem#1\endcsname}%
\let\auto@bib@innerbib\@empty
\bibitem [{\citenamefont {Li}\ \emph {et~al.}(2019)\citenamefont {Li},
  \citenamefont {Lee}, \citenamefont {Wang}, \citenamefont {Osada},
  \citenamefont {Crossley}, \citenamefont {Lee}, \citenamefont {Cui},
  \citenamefont {Hikita},\ and\ \citenamefont {Hwang}}]{D_Li_2019}%
  \BibitemOpen
  \bibfield  {author} {\bibinfo {author} {\bibfnamefont {D.}~\bibnamefont
  {Li}}, \bibinfo {author} {\bibfnamefont {K.}~\bibnamefont {Lee}}, \bibinfo
  {author} {\bibfnamefont {B.~Y.}\ \bibnamefont {Wang}}, \bibinfo {author}
  {\bibfnamefont {M.}~\bibnamefont {Osada}}, \bibinfo {author} {\bibfnamefont
  {S.}~\bibnamefont {Crossley}}, \bibinfo {author} {\bibfnamefont {H.~R.}\
  \bibnamefont {Lee}}, \bibinfo {author} {\bibfnamefont {Y.}~\bibnamefont
  {Cui}}, \bibinfo {author} {\bibfnamefont {Y.}~\bibnamefont {Hikita}},\ and\
  \bibinfo {author} {\bibfnamefont {H.~Y.}\ \bibnamefont {Hwang}},\ }\bibfield
  {title} {\bibinfo {title} {Superconductivity in an infinite-layer
  nickelate},\ }\href {https://doi.org/10.1038/s41586-019-1496-5} {\bibfield
  {journal} {\bibinfo  {journal} {Nature}\ }\textbf {\bibinfo {volume} {572}},\
  \bibinfo {pages} {624} (\bibinfo {year} {2019})}\BibitemShut {NoStop}%
\bibitem [{Y_N()}]{Y_Nomura_2022}%
  \BibitemOpen
  \href@noop {} {}\bibinfo {note} {For a review, see, e.g., Y. Nomura and R.
  Arita, Superconductivity in in^^ef^^ac^^81nite-layer nickelates,
  \href{https://doi.org/10.1088/1361-6633/ac5a60}{Rep. Prog. Phys. \textbf{85},
  052501 (2022)},}\BibitemShut {NoStop}%
\bibitem [{\citenamefont {Chow}\ \emph {et~al.}(2025)\citenamefont {Chow},
  \citenamefont {Luo},\ and\ \citenamefont {Ariando}}]{S_Chow_2025}%
  \BibitemOpen
  \bibfield  {author} {\bibinfo {author} {\bibfnamefont {S.}~\bibnamefont
  {Chow}}, \bibinfo {author} {\bibfnamefont {Z.}~\bibnamefont {Luo}},\ and\
  \bibinfo {author} {\bibfnamefont {A.}~\bibnamefont {Ariando}},\ }\bibfield
  {title} {\bibinfo {title} {Bulk superconductivity near 40 {K} in hole-doped
  \ce{SmNiO2} at ambient pressure},\ }\href
  {https://doi.org/10.1038/s41586-025-08893-4} {\bibfield  {journal} {\bibinfo
  {journal} {Nature}\ }\textbf {\bibinfo {volume} {642}},\ \bibinfo {pages}
  {58} (\bibinfo {year} {2025})}\BibitemShut {NoStop}%
\bibitem [{\citenamefont {Sakakibara}\ \emph {et~al.}(2020)\citenamefont
  {Sakakibara}, \citenamefont {Usui}, \citenamefont {Suzuki}, \citenamefont
  {Kotani}, \citenamefont {Aoki},\ and\ \citenamefont
  {Kuroki}}]{H_Sakakibara_2020}%
  \BibitemOpen
  \bibfield  {author} {\bibinfo {author} {\bibfnamefont {H.}~\bibnamefont
  {Sakakibara}}, \bibinfo {author} {\bibfnamefont {H.}~\bibnamefont {Usui}},
  \bibinfo {author} {\bibfnamefont {K.}~\bibnamefont {Suzuki}}, \bibinfo
  {author} {\bibfnamefont {T.}~\bibnamefont {Kotani}}, \bibinfo {author}
  {\bibfnamefont {H.}~\bibnamefont {Aoki}},\ and\ \bibinfo {author}
  {\bibfnamefont {K.}~\bibnamefont {Kuroki}},\ }\bibfield  {title} {\bibinfo
  {title} {Model construction and a possibility of cupratelike pairing in a new
  ${d}^{9}$ nickelate superconductor \ce{(Nd,Sr)NiO2}},\ }\href
  {https://doi.org/10.1103/PhysRevLett.125.077003} {\bibfield  {journal}
  {\bibinfo  {journal} {Phys. Rev. Lett.}\ }\textbf {\bibinfo {volume} {125}},\
  \bibinfo {pages} {077003} (\bibinfo {year} {2020})}\BibitemShut {NoStop}%
\bibitem [{\citenamefont {Wu}\ \emph {et~al.}(2020)\citenamefont {Wu},
  \citenamefont {Di~Sante}, \citenamefont {Schwemmer}, \citenamefont {Hanke},
  \citenamefont {Hwang}, \citenamefont {Raghu},\ and\ \citenamefont
  {Thomale}}]{X_Wu_2020}%
  \BibitemOpen
  \bibfield  {author} {\bibinfo {author} {\bibfnamefont {X.}~\bibnamefont
  {Wu}}, \bibinfo {author} {\bibfnamefont {D.}~\bibnamefont {Di~Sante}},
  \bibinfo {author} {\bibfnamefont {T.}~\bibnamefont {Schwemmer}}, \bibinfo
  {author} {\bibfnamefont {W.}~\bibnamefont {Hanke}}, \bibinfo {author}
  {\bibfnamefont {H.~Y.}\ \bibnamefont {Hwang}}, \bibinfo {author}
  {\bibfnamefont {S.}~\bibnamefont {Raghu}},\ and\ \bibinfo {author}
  {\bibfnamefont {R.}~\bibnamefont {Thomale}},\ }\bibfield  {title} {\bibinfo
  {title} {Robust ${d}_{{x}^{2}\ensuremath{-}{y}^{2}}$-wave superconductivity
  of infinite-layer nickelates},\ }\href
  {https://doi.org/10.1103/PhysRevB.101.060504} {\bibfield  {journal} {\bibinfo
   {journal} {Phys. Rev. B}\ }\textbf {\bibinfo {volume} {101}},\ \bibinfo
  {pages} {060504} (\bibinfo {year} {2020})}\BibitemShut {NoStop}%
\bibitem [{\citenamefont {Kitatani}\ \emph {et~al.}(2020)\citenamefont
  {Kitatani}, \citenamefont {Si}, \citenamefont {Janson}, \citenamefont
  {Arita}, \citenamefont {Zhong},\ and\ \citenamefont
  {Held}}]{M_Kitatani_2020}%
  \BibitemOpen
  \bibfield  {author} {\bibinfo {author} {\bibfnamefont {M.}~\bibnamefont
  {Kitatani}}, \bibinfo {author} {\bibfnamefont {L.}~\bibnamefont {Si}},
  \bibinfo {author} {\bibfnamefont {O.}~\bibnamefont {Janson}}, \bibinfo
  {author} {\bibfnamefont {R.}~\bibnamefont {Arita}}, \bibinfo {author}
  {\bibfnamefont {Z.}~\bibnamefont {Zhong}},\ and\ \bibinfo {author}
  {\bibfnamefont {K.}~\bibnamefont {Held}},\ }\bibfield  {title} {\bibinfo
  {title} {Nickelate superconductors―a renaissance of the one-band {H}ubbard
  model},\ }\href {https://doi.org/10.1038/s41535-020-00260-y} {\bibfield
  {journal} {\bibinfo  {journal} {npj Quantum Materials}\ }\textbf {\bibinfo
  {volume} {5}},\ \bibinfo {pages} {59} (\bibinfo {year} {2020})}\BibitemShut
  {NoStop}%
\bibitem [{\citenamefont {Kitatani}\ \emph {et~al.}(2023)\citenamefont
  {Kitatani}, \citenamefont {Si}, \citenamefont {Worm}, \citenamefont
  {Tomczak}, \citenamefont {Arita},\ and\ \citenamefont
  {Held}}]{M_Kitatani_2023}%
  \BibitemOpen
  \bibfield  {author} {\bibinfo {author} {\bibfnamefont {M.}~\bibnamefont
  {Kitatani}}, \bibinfo {author} {\bibfnamefont {L.}~\bibnamefont {Si}},
  \bibinfo {author} {\bibfnamefont {P.}~\bibnamefont {Worm}}, \bibinfo {author}
  {\bibfnamefont {J.~M.}\ \bibnamefont {Tomczak}}, \bibinfo {author}
  {\bibfnamefont {R.}~\bibnamefont {Arita}},\ and\ \bibinfo {author}
  {\bibfnamefont {K.}~\bibnamefont {Held}},\ }\bibfield  {title} {\bibinfo
  {title} {Optimizing superconductivity: From cuprates via nickelates to
  palladates},\ }\href {https://doi.org/10.1103/PhysRevLett.130.166002}
  {\bibfield  {journal} {\bibinfo  {journal} {Phys. Rev. Lett.}\ }\textbf
  {\bibinfo {volume} {130}},\ \bibinfo {pages} {166002} (\bibinfo {year}
  {2023})}\BibitemShut {NoStop}%
\bibitem [{\citenamefont {Sun}\ \emph {et~al.}(2023)\citenamefont {Sun},
  \citenamefont {Huo}, \citenamefont {Hu}, \citenamefont {Li}, \citenamefont
  {Liu}, \citenamefont {Han}, \citenamefont {Tang}, \citenamefont {Mao},
  \citenamefont {Yang}, \citenamefont {Wang}, \citenamefont {Cheng},
  \citenamefont {Yao}, \citenamefont {Zhang},\ and\ \citenamefont
  {Wang}}]{H_Sun_2023}%
  \BibitemOpen
  \bibfield  {author} {\bibinfo {author} {\bibfnamefont {H.}~\bibnamefont
  {Sun}}, \bibinfo {author} {\bibfnamefont {M.}~\bibnamefont {Huo}}, \bibinfo
  {author} {\bibfnamefont {X.}~\bibnamefont {Hu}}, \bibinfo {author}
  {\bibfnamefont {J.}~\bibnamefont {Li}}, \bibinfo {author} {\bibfnamefont
  {Z.}~\bibnamefont {Liu}}, \bibinfo {author} {\bibfnamefont {Y.}~\bibnamefont
  {Han}}, \bibinfo {author} {\bibfnamefont {L.}~\bibnamefont {Tang}}, \bibinfo
  {author} {\bibfnamefont {Z.}~\bibnamefont {Mao}}, \bibinfo {author}
  {\bibfnamefont {P.}~\bibnamefont {Yang}}, \bibinfo {author} {\bibfnamefont
  {B.}~\bibnamefont {Wang}}, \bibinfo {author} {\bibfnamefont {J.}~\bibnamefont
  {Cheng}}, \bibinfo {author} {\bibfnamefont {D.-X.}\ \bibnamefont {Yao}},
  \bibinfo {author} {\bibfnamefont {G.-M.}\ \bibnamefont {Zhang}},\ and\
  \bibinfo {author} {\bibfnamefont {M.}~\bibnamefont {Wang}},\ }\bibfield
  {title} {\bibinfo {title} {Signatures of superconductivity near 80 {K} in a
  nickelate under high pressure},\ }\href
  {https://doi.org/10.1038/s41586-023-06408-7} {\bibfield  {journal} {\bibinfo
  {journal} {Nature}\ }\textbf {\bibinfo {volume} {621}},\ \bibinfo {pages}
  {493} (\bibinfo {year} {2023})}\BibitemShut {NoStop}%
\bibitem [{M_W()}]{M_Wang_2024}%
  \BibitemOpen
  \href@noop {} {}\bibinfo {note} {For a review, see, e.g., M. Wang, H.-H. Wen,
  T. Wu, D.-X. Yao, and T. Xiang, Normal and superconducting properties of
  \ce{La3Ni2O7},
  \href{https://www.sciencedirect.com/science/article/pii/S0921453415000477}{Chin.
  Phys. Lett. \textbf{41}, 077402 (2024)},}\BibitemShut {NoStop}%
\bibitem [{\citenamefont {Li}\ \emph {et~al.}(2026)\citenamefont {Li},
  \citenamefont {Xing}, \citenamefont {Peng}, \citenamefont {Dou},
  \citenamefont {Guo}, \citenamefont {Ma}, \citenamefont {Zhang}, \citenamefont
  {Wang}, \citenamefont {Luo}, \citenamefont {Yang}, \citenamefont {Zhang},
  \citenamefont {Chang}, \citenamefont {Chen}, \citenamefont {Cai},
  \citenamefont {Cheng}, \citenamefont {Wang}, \citenamefont {Liu},
  \citenamefont {Luo}, \citenamefont {Hirao}, \citenamefont {Matsuoka},
  \citenamefont {Kadobayashi}, \citenamefont {Zeng}, \citenamefont {Zheng},
  \citenamefont {Zhou}, \citenamefont {Zeng}, \citenamefont {Tao},\ and\
  \citenamefont {Zhang}}]{F_Li_2026}%
  \BibitemOpen
  \bibfield  {author} {\bibinfo {author} {\bibfnamefont {F.}~\bibnamefont
  {Li}}, \bibinfo {author} {\bibfnamefont {Z.}~\bibnamefont {Xing}}, \bibinfo
  {author} {\bibfnamefont {D.}~\bibnamefont {Peng}}, \bibinfo {author}
  {\bibfnamefont {J.}~\bibnamefont {Dou}}, \bibinfo {author} {\bibfnamefont
  {N.}~\bibnamefont {Guo}}, \bibinfo {author} {\bibfnamefont {L.}~\bibnamefont
  {Ma}}, \bibinfo {author} {\bibfnamefont {Y.}~\bibnamefont {Zhang}}, \bibinfo
  {author} {\bibfnamefont {L.}~\bibnamefont {Wang}}, \bibinfo {author}
  {\bibfnamefont {J.}~\bibnamefont {Luo}}, \bibinfo {author} {\bibfnamefont
  {J.}~\bibnamefont {Yang}}, \bibinfo {author} {\bibfnamefont {J.}~\bibnamefont
  {Zhang}}, \bibinfo {author} {\bibfnamefont {T.}~\bibnamefont {Chang}},
  \bibinfo {author} {\bibfnamefont {Y.-S.}\ \bibnamefont {Chen}}, \bibinfo
  {author} {\bibfnamefont {W.}~\bibnamefont {Cai}}, \bibinfo {author}
  {\bibfnamefont {J.}~\bibnamefont {Cheng}}, \bibinfo {author} {\bibfnamefont
  {Y.}~\bibnamefont {Wang}}, \bibinfo {author} {\bibfnamefont {Y.}~\bibnamefont
  {Liu}}, \bibinfo {author} {\bibfnamefont {T.}~\bibnamefont {Luo}}, \bibinfo
  {author} {\bibfnamefont {N.}~\bibnamefont {Hirao}}, \bibinfo {author}
  {\bibfnamefont {T.}~\bibnamefont {Matsuoka}}, \bibinfo {author}
  {\bibfnamefont {H.}~\bibnamefont {Kadobayashi}}, \bibinfo {author}
  {\bibfnamefont {Z.}~\bibnamefont {Zeng}}, \bibinfo {author} {\bibfnamefont
  {Q.}~\bibnamefont {Zheng}}, \bibinfo {author} {\bibfnamefont
  {R.}~\bibnamefont {Zhou}}, \bibinfo {author} {\bibfnamefont {Q.}~\bibnamefont
  {Zeng}}, \bibinfo {author} {\bibfnamefont {X.}~\bibnamefont {Tao}},\ and\
  \bibinfo {author} {\bibfnamefont {J.}~\bibnamefont {Zhang}},\ }\bibfield
  {title} {\bibinfo {title} {Bulk superconductivity up to 96 {K} in pressurized
  nickelate single crystals},\ }\href
  {https://doi.org/10.1038/s41586-025-09954-4} {\bibfield  {journal} {\bibinfo
  {journal} {Nature}\ }\textbf {\bibinfo {volume} {649}},\ \bibinfo {pages}
  {871} (\bibinfo {year} {2026})}\BibitemShut {NoStop}%
\bibitem [{\citenamefont {Ko}\ \emph {et~al.}(2025)\citenamefont {Ko},
  \citenamefont {Yu}, \citenamefont {Liu}, \citenamefont {Bhatt}, \citenamefont
  {Li}, \citenamefont {Thampy}, \citenamefont {Kuo}, \citenamefont {Wang},
  \citenamefont {Lee}, \citenamefont {Lee}, \citenamefont {Lee}, \citenamefont
  {Goodge}, \citenamefont {Muller},\ and\ \citenamefont {Hwang}}]{E_Ko_2025}%
  \BibitemOpen
  \bibfield  {author} {\bibinfo {author} {\bibfnamefont {E.~K.}\ \bibnamefont
  {Ko}}, \bibinfo {author} {\bibfnamefont {Y.}~\bibnamefont {Yu}}, \bibinfo
  {author} {\bibfnamefont {Y.}~\bibnamefont {Liu}}, \bibinfo {author}
  {\bibfnamefont {L.}~\bibnamefont {Bhatt}}, \bibinfo {author} {\bibfnamefont
  {J.}~\bibnamefont {Li}}, \bibinfo {author} {\bibfnamefont {V.}~\bibnamefont
  {Thampy}}, \bibinfo {author} {\bibfnamefont {C.-T.}\ \bibnamefont {Kuo}},
  \bibinfo {author} {\bibfnamefont {B.~Y.}\ \bibnamefont {Wang}}, \bibinfo
  {author} {\bibfnamefont {Y.}~\bibnamefont {Lee}}, \bibinfo {author}
  {\bibfnamefont {K.}~\bibnamefont {Lee}}, \bibinfo {author} {\bibfnamefont
  {J.-S.}\ \bibnamefont {Lee}}, \bibinfo {author} {\bibfnamefont {B.~H.}\
  \bibnamefont {Goodge}}, \bibinfo {author} {\bibfnamefont {D.~A.}\
  \bibnamefont {Muller}},\ and\ \bibinfo {author} {\bibfnamefont {H.~Y.}\
  \bibnamefont {Hwang}},\ }\bibfield  {title} {\bibinfo {title} {Signatures of
  ambient pressure superconductivity in thin film \ce{La3Ni2O7}},\ }\href
  {https://doi.org/10.1038/s41586-024-08525-3} {\bibfield  {journal} {\bibinfo
  {journal} {Nature}\ }\textbf {\bibinfo {volume} {638}},\ \bibinfo {pages}
  {935} (\bibinfo {year} {2025})}\BibitemShut {NoStop}%
\bibitem [{\citenamefont {Zhou}\ \emph
  {et~al.}(2025{\natexlab{a}})\citenamefont {Zhou}, \citenamefont {Lv},
  \citenamefont {Wang}, \citenamefont {Nie}, \citenamefont {Chen},
  \citenamefont {Li}, \citenamefont {Huang}, \citenamefont {Chen},
  \citenamefont {Sun}, \citenamefont {Xue},\ and\ \citenamefont
  {Chen}}]{G_Zhou_2025}%
  \BibitemOpen
  \bibfield  {author} {\bibinfo {author} {\bibfnamefont {G.}~\bibnamefont
  {Zhou}}, \bibinfo {author} {\bibfnamefont {W.}~\bibnamefont {Lv}}, \bibinfo
  {author} {\bibfnamefont {H.}~\bibnamefont {Wang}}, \bibinfo {author}
  {\bibfnamefont {Z.}~\bibnamefont {Nie}}, \bibinfo {author} {\bibfnamefont
  {Y.}~\bibnamefont {Chen}}, \bibinfo {author} {\bibfnamefont {Y.}~\bibnamefont
  {Li}}, \bibinfo {author} {\bibfnamefont {H.}~\bibnamefont {Huang}}, \bibinfo
  {author} {\bibfnamefont {W.-Q.}\ \bibnamefont {Chen}}, \bibinfo {author}
  {\bibfnamefont {Y.-J.}\ \bibnamefont {Sun}}, \bibinfo {author} {\bibfnamefont
  {Q.-K.}\ \bibnamefont {Xue}},\ and\ \bibinfo {author} {\bibfnamefont
  {Z.}~\bibnamefont {Chen}},\ }\bibfield  {title} {\bibinfo {title}
  {Ambient-pressure superconductivity onset above 40{\thinspace}{K} in
  \ce{(La,Pr)3Ni2O7} films},\ }\href
  {https://doi.org/10.1038/s41586-025-08755-z} {\bibfield  {journal} {\bibinfo
  {journal} {Nature}\ }\textbf {\bibinfo {volume} {640}},\ \bibinfo {pages}
  {641} (\bibinfo {year} {2025}{\natexlab{a}})}\BibitemShut {NoStop}%
\bibitem [{\citenamefont {Zhou}\ \emph
  {et~al.}(2025{\natexlab{b}})\citenamefont {Zhou}, \citenamefont {Wang},
  \citenamefont {Huang}, \citenamefont {Chen}, \citenamefont {Peng},
  \citenamefont {Lv}, \citenamefont {Nie}, \citenamefont {Wang}, \citenamefont
  {Xue},\ and\ \citenamefont {Chen}}]{G_Zhou_2025_2}%
  \BibitemOpen
  \bibfield  {author} {\bibinfo {author} {\bibfnamefont {G.}~\bibnamefont
  {Zhou}}, \bibinfo {author} {\bibfnamefont {H.}~\bibnamefont {Wang}}, \bibinfo
  {author} {\bibfnamefont {H.}~\bibnamefont {Huang}}, \bibinfo {author}
  {\bibfnamefont {Y.}~\bibnamefont {Chen}}, \bibinfo {author} {\bibfnamefont
  {F.}~\bibnamefont {Peng}}, \bibinfo {author} {\bibfnamefont {W.}~\bibnamefont
  {Lv}}, \bibinfo {author} {\bibfnamefont {Z.}~\bibnamefont {Nie}}, \bibinfo
  {author} {\bibfnamefont {W.}~\bibnamefont {Wang}}, \bibinfo {author}
  {\bibfnamefont {Q.-K.}\ \bibnamefont {Xue}},\ and\ \bibinfo {author}
  {\bibfnamefont {Z.}~\bibnamefont {Chen}},\ }\href@noop {} {\bibinfo {title}
  {Superconductivity onset above 60 {K} in ambient-pressure nickelate films}}
  (\bibinfo {year} {2025}{\natexlab{b}}),\ \Eprint
  {https://arxiv.org/abs/2512.04708} {arXiv:2512.04708} \BibitemShut {NoStop}%
\bibitem [{\citenamefont {Nakata}\ \emph {et~al.}(2017)\citenamefont {Nakata},
  \citenamefont {Ogura}, \citenamefont {Usui},\ and\ \citenamefont
  {Kuroki}}]{M_Nakata_2017}%
  \BibitemOpen
  \bibfield  {author} {\bibinfo {author} {\bibfnamefont {M.}~\bibnamefont
  {Nakata}}, \bibinfo {author} {\bibfnamefont {D.}~\bibnamefont {Ogura}},
  \bibinfo {author} {\bibfnamefont {H.}~\bibnamefont {Usui}},\ and\ \bibinfo
  {author} {\bibfnamefont {K.}~\bibnamefont {Kuroki}},\ }\bibfield  {title}
  {\bibinfo {title} {Finite-energy spin fluctuations as a pairing glue in
  systems with coexisting electron and hole bands},\ }\href
  {https://link.aps.org/doi/10.1103/PhysRevB.95.214509} {\bibfield  {journal}
  {\bibinfo  {journal} {Phys. Rev. B}\ }\textbf {\bibinfo {volume} {95}},\
  \bibinfo {pages} {214509} (\bibinfo {year} {2017})}\BibitemShut {NoStop}%
\bibitem [{\citenamefont {Sakakibara}\ \emph {et~al.}(2010)\citenamefont
  {Sakakibara}, \citenamefont {Usui}, \citenamefont {Kuroki}, \citenamefont
  {Arita},\ and\ \citenamefont {Aoki}}]{H_Sakakibara_2010}%
  \BibitemOpen
  \bibfield  {author} {\bibinfo {author} {\bibfnamefont {H.}~\bibnamefont
  {Sakakibara}}, \bibinfo {author} {\bibfnamefont {H.}~\bibnamefont {Usui}},
  \bibinfo {author} {\bibfnamefont {K.}~\bibnamefont {Kuroki}}, \bibinfo
  {author} {\bibfnamefont {R.}~\bibnamefont {Arita}},\ and\ \bibinfo {author}
  {\bibfnamefont {H.}~\bibnamefont {Aoki}},\ }\bibfield  {title} {\bibinfo
  {title} {Two-orbital model explains the higher transition temperature of the
  single-layer {H}g-cuprate superconductor compared to that of the {L}a-cuprate
  superconductor},\ }\href {https://doi.org/10.1103/PhysRevLett.105.057003}
  {\bibfield  {journal} {\bibinfo  {journal} {Phys. Rev. Lett.}\ }\textbf
  {\bibinfo {volume} {105}},\ \bibinfo {pages} {057003} (\bibinfo {year}
  {2010})}\BibitemShut {NoStop}%
\bibitem [{\citenamefont {Sakakibara}\ \emph {et~al.}(2012)\citenamefont
  {Sakakibara}, \citenamefont {Suzuki}, \citenamefont {Usui}, \citenamefont
  {Kuroki}, \citenamefont {Arita}, \citenamefont {Scalapino},\ and\
  \citenamefont {Aoki}}]{H_Sakakibara_2012}%
  \BibitemOpen
  \bibfield  {author} {\bibinfo {author} {\bibfnamefont {H.}~\bibnamefont
  {Sakakibara}}, \bibinfo {author} {\bibfnamefont {K.}~\bibnamefont {Suzuki}},
  \bibinfo {author} {\bibfnamefont {H.}~\bibnamefont {Usui}}, \bibinfo {author}
  {\bibfnamefont {K.}~\bibnamefont {Kuroki}}, \bibinfo {author} {\bibfnamefont
  {R.}~\bibnamefont {Arita}}, \bibinfo {author} {\bibfnamefont {D.~J.}\
  \bibnamefont {Scalapino}},\ and\ \bibinfo {author} {\bibfnamefont
  {H.}~\bibnamefont {Aoki}},\ }\bibfield  {title} {\bibinfo {title}
  {Multiorbital analysis of the effects of uniaxial and hydrostatic pressure on
  ${T}_{c}$ in the single-layered cuprate superconductors},\ }\href
  {https://doi.org/10.1103/PhysRevB.86.134520} {\bibfield  {journal} {\bibinfo
  {journal} {Phys. Rev. B}\ }\textbf {\bibinfo {volume} {86}},\ \bibinfo
  {pages} {134520} (\bibinfo {year} {2012})}\BibitemShut {NoStop}%
\bibitem [{\citenamefont {Monthoux}\ and\ \citenamefont
  {Lonzarich}(1999)}]{P_Monthoux_1999}%
  \BibitemOpen
  \bibfield  {author} {\bibinfo {author} {\bibfnamefont {P.}~\bibnamefont
  {Monthoux}}\ and\ \bibinfo {author} {\bibfnamefont {G.~G.}\ \bibnamefont
  {Lonzarich}},\ }\bibfield  {title} {\bibinfo {title} {$\mathit{p}$-wave and
  $\mathit{d}$-wave superconductivity in quasi-two-dimensional metals},\ }\href
  {https://doi.org/10.1103/PhysRevB.59.14598} {\bibfield  {journal} {\bibinfo
  {journal} {Phys. Rev. B}\ }\textbf {\bibinfo {volume} {59}},\ \bibinfo
  {pages} {14598} (\bibinfo {year} {1999})}\BibitemShut {NoStop}%
\bibitem [{\citenamefont {Arita}\ \emph {et~al.}(2000)\citenamefont {Arita},
  \citenamefont {Kuroki},\ and\ \citenamefont {Aoki}}]{R_Arita_2000}%
  \BibitemOpen
  \bibfield  {author} {\bibinfo {author} {\bibfnamefont {R.}~\bibnamefont
  {Arita}}, \bibinfo {author} {\bibfnamefont {K.}~\bibnamefont {Kuroki}},\ and\
  \bibinfo {author} {\bibfnamefont {H.}~\bibnamefont {Aoki}},\ }\bibfield
  {title} {\bibinfo {title} {d- and p-wave superconductivity mediated by spin
  fluctuations in two- and three-dimensional single-band repulsive hubbard
  model},\ }\href {https://doi.org/10.1143/JPSJ.69.1181} {\bibfield  {journal}
  {\bibinfo  {journal} {J. Phys. Soc. Jpn.}\ }\textbf {\bibinfo {volume}
  {69}},\ \bibinfo {pages} {1181} (\bibinfo {year} {2000})}\BibitemShut
  {NoStop}%
\bibitem [{\citenamefont {Arita}\ \emph {et~al.}(2001)\citenamefont {Arita},
  \citenamefont {Kuroki},\ and\ \citenamefont {Aoki}}]{R_Arita_2001}%
  \BibitemOpen
  \bibfield  {author} {\bibinfo {author} {\bibfnamefont {R.}~\bibnamefont
  {Arita}}, \bibinfo {author} {\bibfnamefont {K.}~\bibnamefont {Kuroki}},\ and\
  \bibinfo {author} {\bibfnamefont {H.}~\bibnamefont {Aoki}},\ }\bibfield
  {title} {\bibinfo {title} {Fluctuation exchange study of singlet and triplet
  superconductivity in 2{D} and 3{D} single-band {H}ubbard model},\ }\href
  {https://doi.org/https://doi.org/10.1016/S0022-3697(00)00138-4} {\bibfield
  {journal} {\bibinfo  {journal} {J. Phys. Chem. Solids}\ }\textbf {\bibinfo
  {volume} {62}},\ \bibinfo {pages} {249} (\bibinfo {year} {2001})}\BibitemShut
  {NoStop}%
\bibitem [{\citenamefont {Dagotto}\ \emph {et~al.}(1992)\citenamefont
  {Dagotto}, \citenamefont {Riera},\ and\ \citenamefont
  {Scalapino}}]{bilayer0}%
  \BibitemOpen
  \bibfield  {author} {\bibinfo {author} {\bibfnamefont {E.}~\bibnamefont
  {Dagotto}}, \bibinfo {author} {\bibfnamefont {J.}~\bibnamefont {Riera}},\
  and\ \bibinfo {author} {\bibfnamefont {D.}~\bibnamefont {Scalapino}},\
  }\bibfield  {title} {\bibinfo {title} {Superconductivity in ladders and
  coupled planes},\ }\href {https://doi.org/10.1103/PhysRevB.45.5744}
  {\bibfield  {journal} {\bibinfo  {journal} {Phys. Rev. B}\ }\textbf {\bibinfo
  {volume} {45}},\ \bibinfo {pages} {5744} (\bibinfo {year}
  {1992})}\BibitemShut {NoStop}%
\bibitem [{\citenamefont {Bulut}\ \emph {et~al.}(1992)\citenamefont {Bulut},
  \citenamefont {Scalapino},\ and\ \citenamefont {Scalettar}}]{bilayer1}%
  \BibitemOpen
  \bibfield  {author} {\bibinfo {author} {\bibfnamefont {N.}~\bibnamefont
  {Bulut}}, \bibinfo {author} {\bibfnamefont {D.~J.}\ \bibnamefont
  {Scalapino}},\ and\ \bibinfo {author} {\bibfnamefont {R.~T.}\ \bibnamefont
  {Scalettar}},\ }\bibfield  {title} {\bibinfo {title} {Nodeless d-wave pairing
  in a two-layer hubbard model},\ }\href
  {https://doi.org/10.1103/PhysRevB.45.5577} {\bibfield  {journal} {\bibinfo
  {journal} {Phys. Rev. B}\ }\textbf {\bibinfo {volume} {45}},\ \bibinfo
  {pages} {5577} (\bibinfo {year} {1992})}\BibitemShut {NoStop}%
\bibitem [{\citenamefont {Scalettar}\ \emph {et~al.}(1994)\citenamefont
  {Scalettar}, \citenamefont {Cannon}, \citenamefont {Scalapino},\ and\
  \citenamefont {Sugar}}]{bilayer2}%
  \BibitemOpen
  \bibfield  {author} {\bibinfo {author} {\bibfnamefont {R.~T.}\ \bibnamefont
  {Scalettar}}, \bibinfo {author} {\bibfnamefont {J.~W.}\ \bibnamefont
  {Cannon}}, \bibinfo {author} {\bibfnamefont {D.~J.}\ \bibnamefont
  {Scalapino}},\ and\ \bibinfo {author} {\bibfnamefont {R.~L.}\ \bibnamefont
  {Sugar}},\ }\bibfield  {title} {\bibinfo {title} {Magnetic and pairing
  correlations in coupled hubbard planes},\ }\href
  {https://doi.org/10.1103/PhysRevB.50.13419} {\bibfield  {journal} {\bibinfo
  {journal} {Phys. Rev. B}\ }\textbf {\bibinfo {volume} {50}},\ \bibinfo
  {pages} {13419} (\bibinfo {year} {1994})}\BibitemShut {NoStop}%
\bibitem [{\citenamefont {Hetzel}\ \emph {et~al.}(1994)\citenamefont {Hetzel},
  \citenamefont {von~der Linden},\ and\ \citenamefont {Hanke}}]{bilayer3}%
  \BibitemOpen
  \bibfield  {author} {\bibinfo {author} {\bibfnamefont {R.~E.}\ \bibnamefont
  {Hetzel}}, \bibinfo {author} {\bibfnamefont {W.}~\bibnamefont {von~der
  Linden}},\ and\ \bibinfo {author} {\bibfnamefont {W.}~\bibnamefont {Hanke}},\
  }\bibfield  {title} {\bibinfo {title} {Pairing correlations in a two-layer
  hubbard model},\ }\href {https://doi.org/10.1103/PhysRevB.50.4159} {\bibfield
   {journal} {\bibinfo  {journal} {Phys. Rev. B}\ }\textbf {\bibinfo {volume}
  {50}},\ \bibinfo {pages} {4159} (\bibinfo {year} {1994})}\BibitemShut
  {NoStop}%
\bibitem [{\citenamefont {dos Santos}(1995)}]{bilayer4}%
  \BibitemOpen
  \bibfield  {author} {\bibinfo {author} {\bibfnamefont {R.~R.}\ \bibnamefont
  {dos Santos}},\ }\bibfield  {title} {\bibinfo {title} {Magnetism and pairing
  in hubbard bilayers},\ }\href {https://doi.org/10.1103/PhysRevB.51.15540}
  {\bibfield  {journal} {\bibinfo  {journal} {Phys. Rev. B}\ }\textbf {\bibinfo
  {volume} {51}},\ \bibinfo {pages} {15540} (\bibinfo {year}
  {1995})}\BibitemShut {NoStop}%
\bibitem [{\citenamefont {Liechtenstein}\ \emph {et~al.}(1995)\citenamefont
  {Liechtenstein}, \citenamefont {Mazin},\ and\ \citenamefont
  {Andersen}}]{bilayer5}%
  \BibitemOpen
  \bibfield  {author} {\bibinfo {author} {\bibfnamefont {A.~I.}\ \bibnamefont
  {Liechtenstein}}, \bibinfo {author} {\bibfnamefont {I.~I.}\ \bibnamefont
  {Mazin}},\ and\ \bibinfo {author} {\bibfnamefont {O.~K.}\ \bibnamefont
  {Andersen}},\ }\bibfield  {title} {\bibinfo {title} {$\mathit{s}$-wave
  superconductivity from an antiferromagnetic spin-fluctuation model for
  bilayer materials},\ }\href {https://doi.org/10.1103/PhysRevLett.74.2303}
  {\bibfield  {journal} {\bibinfo  {journal} {Phys. Rev. Lett.}\ }\textbf
  {\bibinfo {volume} {74}},\ \bibinfo {pages} {2303} (\bibinfo {year}
  {1995})}\BibitemShut {NoStop}%
\bibitem [{\citenamefont {Kuroki}\ \emph {et~al.}(2002)\citenamefont {Kuroki},
  \citenamefont {Kimura},\ and\ \citenamefont {Arita}}]{bilayer6}%
  \BibitemOpen
  \bibfield  {author} {\bibinfo {author} {\bibfnamefont {K.}~\bibnamefont
  {Kuroki}}, \bibinfo {author} {\bibfnamefont {T.}~\bibnamefont {Kimura}},\
  and\ \bibinfo {author} {\bibfnamefont {R.}~\bibnamefont {Arita}},\ }\bibfield
   {title} {\bibinfo {title} {High-temperature superconductivity in dimer array
  systems},\ }\href {https://doi.org/10.1103/PhysRevB.66.184508} {\bibfield
  {journal} {\bibinfo  {journal} {Phys. Rev. B}\ }\textbf {\bibinfo {volume}
  {66}},\ \bibinfo {pages} {184508} (\bibinfo {year} {2002})}\BibitemShut
  {NoStop}%
\bibitem [{\citenamefont {Kancharla}\ and\ \citenamefont
  {Okamoto}(2007)}]{bilayer7}%
  \BibitemOpen
  \bibfield  {author} {\bibinfo {author} {\bibfnamefont {S.~S.}\ \bibnamefont
  {Kancharla}}\ and\ \bibinfo {author} {\bibfnamefont {S.}~\bibnamefont
  {Okamoto}},\ }\bibfield  {title} {\bibinfo {title} {Band insulator to mott
  insulator transition in a bilayer hubbard model},\ }\href
  {https://doi.org/10.1103/PhysRevB.75.193103} {\bibfield  {journal} {\bibinfo
  {journal} {Phys. Rev. B}\ }\textbf {\bibinfo {volume} {75}},\ \bibinfo
  {pages} {193103} (\bibinfo {year} {2007})}\BibitemShut {NoStop}%
\bibitem [{\citenamefont {Bouadim}\ \emph {et~al.}(2008)\citenamefont
  {Bouadim}, \citenamefont {Batrouni}, \citenamefont {H\'{e}bert},\ and\
  \citenamefont {Scalettar}}]{bilayer8}%
  \BibitemOpen
  \bibfield  {author} {\bibinfo {author} {\bibfnamefont {K.}~\bibnamefont
  {Bouadim}}, \bibinfo {author} {\bibfnamefont {G.~G.}\ \bibnamefont
  {Batrouni}}, \bibinfo {author} {\bibfnamefont {F.}~\bibnamefont
  {H\'{e}bert}},\ and\ \bibinfo {author} {\bibfnamefont {R.~T.}\ \bibnamefont
  {Scalettar}},\ }\bibfield  {title} {\bibinfo {title} {Magnetic and transport
  properties of a coupled hubbard bilayer with electron and hole doping},\
  }\href {https://doi.org/10.1103/PhysRevB.77.144527} {\bibfield  {journal}
  {\bibinfo  {journal} {Phys. Rev. B}\ }\textbf {\bibinfo {volume} {77}},\
  \bibinfo {pages} {144527} (\bibinfo {year} {2008})}\BibitemShut {NoStop}%
\bibitem [{\citenamefont {Lanat\`a}\ \emph {et~al.}(2009)\citenamefont
  {Lanat\`a}, \citenamefont {Barone},\ and\ \citenamefont
  {Fabrizio}}]{bilayer9}%
  \BibitemOpen
  \bibfield  {author} {\bibinfo {author} {\bibfnamefont {N.}~\bibnamefont
  {Lanat\`a}}, \bibinfo {author} {\bibfnamefont {P.}~\bibnamefont {Barone}},\
  and\ \bibinfo {author} {\bibfnamefont {M.}~\bibnamefont {Fabrizio}},\
  }\bibfield  {title} {\bibinfo {title} {Superconductivity in the doped bilayer
  hubbard model},\ }\href {https://doi.org/10.1103/PhysRevB.80.224524}
  {\bibfield  {journal} {\bibinfo  {journal} {Phys. Rev. B}\ }\textbf {\bibinfo
  {volume} {80}},\ \bibinfo {pages} {224524} (\bibinfo {year}
  {2009})}\BibitemShut {NoStop}%
\bibitem [{\citenamefont {Zhai}\ \emph {et~al.}(2009)\citenamefont {Zhai},
  \citenamefont {Wang},\ and\ \citenamefont {Lee}}]{bilayer10}%
  \BibitemOpen
  \bibfield  {author} {\bibinfo {author} {\bibfnamefont {H.}~\bibnamefont
  {Zhai}}, \bibinfo {author} {\bibfnamefont {F.}~\bibnamefont {Wang}},\ and\
  \bibinfo {author} {\bibfnamefont {D.-H.}\ \bibnamefont {Lee}},\ }\bibfield
  {title} {\bibinfo {title} {Antiferromagnetically driven electronic
  correlations in iron pnictides and cuprates},\ }\href
  {https://doi.org/10.1103/PhysRevB.80.064517} {\bibfield  {journal} {\bibinfo
  {journal} {Phys. Rev. B}\ }\textbf {\bibinfo {volume} {80}},\ \bibinfo
  {pages} {064517} (\bibinfo {year} {2009})}\BibitemShut {NoStop}%
\bibitem [{\citenamefont {Maier}\ and\ \citenamefont
  {Scalapino}(2011)}]{bilayer11}%
  \BibitemOpen
  \bibfield  {author} {\bibinfo {author} {\bibfnamefont {T.~A.}\ \bibnamefont
  {Maier}}\ and\ \bibinfo {author} {\bibfnamefont {D.~J.}\ \bibnamefont
  {Scalapino}},\ }\bibfield  {title} {\bibinfo {title} {Pair structure and the
  pairing interaction in a bilayer hubbard model for unconventional
  superconductivity},\ }\href {https://doi.org/10.1103/PhysRevB.84.180513}
  {\bibfield  {journal} {\bibinfo  {journal} {Phys. Rev. B}\ }\textbf {\bibinfo
  {volume} {84}},\ \bibinfo {pages} {180513} (\bibinfo {year}
  {2011})}\BibitemShut {NoStop}%
\bibitem [{\citenamefont {Nomura}\ \emph {et~al.}(2025)\citenamefont {Nomura},
  \citenamefont {Kitatani}, \citenamefont {Sakai},\ and\ \citenamefont
  {Arita}}]{Y_Nomura_2025}%
  \BibitemOpen
  \bibfield  {author} {\bibinfo {author} {\bibfnamefont {Y.}~\bibnamefont
  {Nomura}}, \bibinfo {author} {\bibfnamefont {M.}~\bibnamefont {Kitatani}},
  \bibinfo {author} {\bibfnamefont {S.}~\bibnamefont {Sakai}},\ and\ \bibinfo
  {author} {\bibfnamefont {R.}~\bibnamefont {Arita}},\ }\bibfield  {title}
  {\bibinfo {title} {Strong-coupling high-${T}_{c}$ superconductivity in doped
  correlated band insulators},\ }\href {https://doi.org/10.1103/dygc-94fq}
  {\bibfield  {journal} {\bibinfo  {journal} {Phys. Rev. B}\ }\textbf {\bibinfo
  {volume} {112}},\ \bibinfo {pages} {L020504} (\bibinfo {year}
  {2025})}\BibitemShut {NoStop}%
\bibitem [{\citenamefont {Hirschfeld}\ \emph {et~al.}(2011)\citenamefont
  {Hirschfeld}, \citenamefont {Korshunov},\ and\ \citenamefont
  {Mazin}}]{P_Hirschfeld_2011}%
  \BibitemOpen
  \bibfield  {author} {\bibinfo {author} {\bibfnamefont {P.~J.}\ \bibnamefont
  {Hirschfeld}}, \bibinfo {author} {\bibfnamefont {M.~M.}\ \bibnamefont
  {Korshunov}},\ and\ \bibinfo {author} {\bibfnamefont {I.~I.}\ \bibnamefont
  {Mazin}},\ }\bibfield  {title} {\bibinfo {title} {Gap symmetry and structure
  of fe-based superconductors},\ }\href
  {https://doi.org/10.1088/0034-4885/74/12/124508} {\bibfield  {journal}
  {\bibinfo  {journal} {Rep. Prog. Phys.}\ }\textbf {\bibinfo {volume} {74}},\
  \bibinfo {pages} {124508} (\bibinfo {year} {2011})}\BibitemShut {NoStop}%
\bibitem [{\citenamefont {Bang}(2016)}]{Y_Bang_2016}%
  \BibitemOpen
  \bibfield  {author} {\bibinfo {author} {\bibfnamefont {Y.}~\bibnamefont
  {Bang}},\ }\bibfield  {title} {\bibinfo {title} {Pairing mechanism of heavily
  electron doped fese systems: dynamical tuning of the pairing cutoff energy},\
  }\href {https://doi.org/10.1088/1367-2630/18/11/113054} {\bibfield  {journal}
  {\bibinfo  {journal} {New J. Phys.}\ }\textbf {\bibinfo {volume} {18}},\
  \bibinfo {pages} {113054} (\bibinfo {year} {2016})}\BibitemShut {NoStop}%
\bibitem [{\citenamefont {Bang}(2019)}]{Y_Bang_2019}%
  \BibitemOpen
  \bibfield  {author} {\bibinfo {author} {\bibfnamefont {Y.}~\bibnamefont
  {Bang}},\ }\bibfield  {title} {\bibinfo {title} {Phonon boost effect on the
  $s\pm$-wave superconductor with incipient band},\ }\href
  {https://doi.org/10.1038/s41598-019-40536-3} {\bibfield  {journal} {\bibinfo
  {journal} {Sci. Rep.}\ }\textbf {\bibinfo {volume} {9}},\ \bibinfo {pages}
  {3907} (\bibinfo {year} {2019})}\BibitemShut {NoStop}%
\bibitem [{\citenamefont {Kuroki}\ \emph {et~al.}(2005)\citenamefont {Kuroki},
  \citenamefont {Higashida},\ and\ \citenamefont {Arita}}]{K_Kuroki_2005}%
  \BibitemOpen
  \bibfield  {author} {\bibinfo {author} {\bibfnamefont {K.}~\bibnamefont
  {Kuroki}}, \bibinfo {author} {\bibfnamefont {T.}~\bibnamefont {Higashida}},\
  and\ \bibinfo {author} {\bibfnamefont {R.}~\bibnamefont {Arita}},\ }\bibfield
   {title} {\bibinfo {title} {High-${T}_{c}$ superconductivity due to
  coexisting wide and narrow bands: A fluctuation exchange study of the hubbard
  ladder as a test case},\ }\href {https://doi.org/10.1103/PhysRevB.72.212509}
  {\bibfield  {journal} {\bibinfo  {journal} {Phys. Rev. B}\ }\textbf {\bibinfo
  {volume} {72}},\ \bibinfo {pages} {212509} (\bibinfo {year}
  {2005})}\BibitemShut {NoStop}%
\bibitem [{\citenamefont {Ogura}\ \emph {et~al.}(2017)\citenamefont {Ogura},
  \citenamefont {Aoki},\ and\ \citenamefont {Kuroki}}]{D_Ogura_2017}%
  \BibitemOpen
  \bibfield  {author} {\bibinfo {author} {\bibfnamefont {D.}~\bibnamefont
  {Ogura}}, \bibinfo {author} {\bibfnamefont {H.}~\bibnamefont {Aoki}},\ and\
  \bibinfo {author} {\bibfnamefont {K.}~\bibnamefont {Kuroki}},\ }\bibfield
  {title} {\bibinfo {title} {Possible high-${T}_{c}$ superconductivity due to
  incipient narrow bands originating from hidden ladders in ruddlesden-popper
  compounds},\ }\href {https://doi.org/10.1103/PhysRevB.96.184513} {\bibfield
  {journal} {\bibinfo  {journal} {Phys. Rev. B}\ }\textbf {\bibinfo {volume}
  {96}},\ \bibinfo {pages} {184513} (\bibinfo {year} {2017})}\BibitemShut
  {NoStop}%
\bibitem [{\citenamefont {Matsumoto}\ \emph {et~al.}(2018)\citenamefont
  {Matsumoto}, \citenamefont {Ogura},\ and\ \citenamefont
  {Kuroki}}]{K_Matsumoto_2018}%
  \BibitemOpen
  \bibfield  {author} {\bibinfo {author} {\bibfnamefont {K.}~\bibnamefont
  {Matsumoto}}, \bibinfo {author} {\bibfnamefont {D.}~\bibnamefont {Ogura}},\
  and\ \bibinfo {author} {\bibfnamefont {K.}~\bibnamefont {Kuroki}},\
  }\bibfield  {title} {\bibinfo {title} {Wide applicability of high-${T}_{c}$
  pairing originating from coexisting wide and incipient narrow bands in
  quasi-one-dimensional systems},\ }\href
  {https://doi.org/10.1103/PhysRevB.97.014516} {\bibfield  {journal} {\bibinfo
  {journal} {Phys. Rev. B}\ }\textbf {\bibinfo {volume} {97}},\ \bibinfo
  {pages} {014516} (\bibinfo {year} {2018})}\BibitemShut {NoStop}%
\bibitem [{\citenamefont {Matsumoto}\ \emph {et~al.}(2020)\citenamefont
  {Matsumoto}, \citenamefont {Ogura},\ and\ \citenamefont
  {Kuroki}}]{K_Matsumoto_2020}%
  \BibitemOpen
  \bibfield  {author} {\bibinfo {author} {\bibfnamefont {K.}~\bibnamefont
  {Matsumoto}}, \bibinfo {author} {\bibfnamefont {D.}~\bibnamefont {Ogura}},\
  and\ \bibinfo {author} {\bibfnamefont {K.}~\bibnamefont {Kuroki}},\
  }\bibfield  {title} {\bibinfo {title} {Strongly enhanced superconductivity
  due to finite energy spin fluctuations induced by an incipient band: A flex
  study on the bilayer hubbard model with vertical and diagonal interlayer
  hoppings},\ }\href {https://doi.org/10.7566/JPSJ.89.044709} {\bibfield
  {journal} {\bibinfo  {journal} {J. Phys. Soc. Jpn.}\ }\textbf {\bibinfo
  {volume} {89}},\ \bibinfo {pages} {044709} (\bibinfo {year}
  {2020})}\BibitemShut {NoStop}%
\bibitem [{\citenamefont {Sakamoto}\ and\ \citenamefont
  {Kuroki}(2020)}]{H_Sakamoto_2020}%
  \BibitemOpen
  \bibfield  {author} {\bibinfo {author} {\bibfnamefont {H.}~\bibnamefont
  {Sakamoto}}\ and\ \bibinfo {author} {\bibfnamefont {K.}~\bibnamefont
  {Kuroki}},\ }\bibfield  {title} {\bibinfo {title} {Possible enhancement of
  superconductivity in ladder-type cuprates by longitudinal compression},\
  }\href {https://doi.org/10.1103/PhysRevResearch.2.022055} {\bibfield
  {journal} {\bibinfo  {journal} {Phys. Rev. Res.}\ }\textbf {\bibinfo {volume}
  {2}},\ \bibinfo {pages} {022055} (\bibinfo {year} {2020})}\BibitemShut
  {NoStop}%
\bibitem [{\citenamefont {Kato}\ and\ \citenamefont
  {Kuroki}(2020)}]{D_Kato_2020}%
  \BibitemOpen
  \bibfield  {author} {\bibinfo {author} {\bibfnamefont {D.}~\bibnamefont
  {Kato}}\ and\ \bibinfo {author} {\bibfnamefont {K.}~\bibnamefont {Kuroki}},\
  }\bibfield  {title} {\bibinfo {title} {Many-variable variational monte carlo
  study of superconductivity in two-band hubbard models with an incipient
  band},\ }\href {https://doi.org/10.1103/PhysRevResearch.2.023156} {\bibfield
  {journal} {\bibinfo  {journal} {Phys. Rev. Res.}\ }\textbf {\bibinfo {volume}
  {2}},\ \bibinfo {pages} {023156} (\bibinfo {year} {2020})}\BibitemShut
  {NoStop}%
\bibitem [{\citenamefont {Aida}\ \emph {et~al.}(2024)\citenamefont {Aida},
  \citenamefont {Matsumoto}, \citenamefont {Ogura}, \citenamefont {Ochi},\ and\
  \citenamefont {Kuroki}}]{T_Aida_2024}%
  \BibitemOpen
  \bibfield  {author} {\bibinfo {author} {\bibfnamefont {T.}~\bibnamefont
  {Aida}}, \bibinfo {author} {\bibfnamefont {K.}~\bibnamefont {Matsumoto}},
  \bibinfo {author} {\bibfnamefont {D.}~\bibnamefont {Ogura}}, \bibinfo
  {author} {\bibfnamefont {M.}~\bibnamefont {Ochi}},\ and\ \bibinfo {author}
  {\bibfnamefont {K.}~\bibnamefont {Kuroki}},\ }\bibfield  {title} {\bibinfo
  {title} {Theoretical study of spin-fluctuation-mediated superconductivity in
  two-dimensional hubbard models with an incipient flat band},\ }\href
  {https://doi.org/10.1103/PhysRevB.110.054516} {\bibfield  {journal} {\bibinfo
   {journal} {Phys. Rev. B}\ }\textbf {\bibinfo {volume} {110}},\ \bibinfo
  {pages} {054516} (\bibinfo {year} {2024})}\BibitemShut {NoStop}%
\bibitem [{\citenamefont {Yagi}\ \emph {et~al.}(2024)\citenamefont {Yagi},
  \citenamefont {Ochi},\ and\ \citenamefont {Kuroki}}]{T_Yagi_2024}%
  \BibitemOpen
  \bibfield  {author} {\bibinfo {author} {\bibfnamefont {T.}~\bibnamefont
  {Yagi}}, \bibinfo {author} {\bibfnamefont {M.}~\bibnamefont {Ochi}},\ and\
  \bibinfo {author} {\bibfnamefont {K.}~\bibnamefont {Kuroki}},\ }\bibfield
  {title} {\bibinfo {title} {Theoretical analysis of the origin of the
  double-well band dispersion in the cuo double chains of
  \ce{Pr2Ba4Cu7O_{15-$\delta$}} and its impact on superconductivity},\ }\href
  {https://doi.org/10.1103/PhysRevB.110.184516} {\bibfield  {journal} {\bibinfo
   {journal} {Phys. Rev. B}\ }\textbf {\bibinfo {volume} {110}},\ \bibinfo
  {pages} {184516} (\bibinfo {year} {2024})}\BibitemShut {NoStop}%
\bibitem [{\citenamefont {Yamazaki}\ \emph {et~al.}(2020)\citenamefont
  {Yamazaki}, \citenamefont {Ochi}, \citenamefont {Ogura}, \citenamefont
  {Kuroki}, \citenamefont {Eisaki}, \citenamefont {Uchida},\ and\ \citenamefont
  {Aoki}}]{K_Yamazaki_2020}%
  \BibitemOpen
  \bibfield  {author} {\bibinfo {author} {\bibfnamefont {K.}~\bibnamefont
  {Yamazaki}}, \bibinfo {author} {\bibfnamefont {M.}~\bibnamefont {Ochi}},
  \bibinfo {author} {\bibfnamefont {D.}~\bibnamefont {Ogura}}, \bibinfo
  {author} {\bibfnamefont {K.}~\bibnamefont {Kuroki}}, \bibinfo {author}
  {\bibfnamefont {H.}~\bibnamefont {Eisaki}}, \bibinfo {author} {\bibfnamefont
  {S.}~\bibnamefont {Uchida}},\ and\ \bibinfo {author} {\bibfnamefont
  {H.}~\bibnamefont {Aoki}},\ }\bibfield  {title} {\bibinfo {title}
  {Superconducting mechanism for the cuprate \ce{Ba2Cu3O_{3+$\delta$}} based on
  a multiorbital lieb lattice model},\ }\href
  {https://doi.org/10.1103/PhysRevResearch.2.033356} {\bibfield  {journal}
  {\bibinfo  {journal} {Phys. Rev. Res.}\ }\textbf {\bibinfo {volume} {2}},\
  \bibinfo {pages} {033356} (\bibinfo {year} {2020})}\BibitemShut {NoStop}%
\bibitem [{\citenamefont {Kitamine}\ \emph {et~al.}(2020)\citenamefont
  {Kitamine}, \citenamefont {Ochi},\ and\ \citenamefont
  {Kuroki}}]{N_Kitamine_2020}%
  \BibitemOpen
  \bibfield  {author} {\bibinfo {author} {\bibfnamefont {N.}~\bibnamefont
  {Kitamine}}, \bibinfo {author} {\bibfnamefont {M.}~\bibnamefont {Ochi}},\
  and\ \bibinfo {author} {\bibfnamefont {K.}~\bibnamefont {Kuroki}},\
  }\bibfield  {title} {\bibinfo {title} {Designing nickelate superconductors
  with ${d}^{8}$ configuration exploiting mixed-anion strategy},\ }\href
  {https://doi.org/10.1103/PhysRevResearch.2.042032} {\bibfield  {journal}
  {\bibinfo  {journal} {Phys. Rev. Res.}\ }\textbf {\bibinfo {volume} {2}},\
  \bibinfo {pages} {042032} (\bibinfo {year} {2020})}\BibitemShut {NoStop}%
\bibitem [{\citenamefont {Kitamine}\ \emph {et~al.}(2023)\citenamefont
  {Kitamine}, \citenamefont {Ochi},\ and\ \citenamefont
  {Kuroki}}]{N_Kitamine_2023}%
  \BibitemOpen
  \bibfield  {author} {\bibinfo {author} {\bibfnamefont {N.}~\bibnamefont
  {Kitamine}}, \bibinfo {author} {\bibfnamefont {M.}~\bibnamefont {Ochi}},\
  and\ \bibinfo {author} {\bibfnamefont {K.}~\bibnamefont {Kuroki}},\
  }\href@noop {} {\bibinfo {title} {Theoretical designing of multiband
  nickelate and palladate superconductors with $d^{8+\delta}$ configuration}}
  (\bibinfo {year} {2023}),\ \Eprint {https://arxiv.org/abs/2308.12750}
  {arXiv:2308.12750} \BibitemShut {NoStop}%
\bibitem [{\citenamefont {Sakakibara}\ \emph {et~al.}(2025)\citenamefont
  {Sakakibara}, \citenamefont {Mizuno}, \citenamefont {Ochi}, \citenamefont
  {Usui},\ and\ \citenamefont {Kuroki}}]{H_Sakakibara_2025}%
  \BibitemOpen
  \bibfield  {author} {\bibinfo {author} {\bibfnamefont {H.}~\bibnamefont
  {Sakakibara}}, \bibinfo {author} {\bibfnamefont {R.}~\bibnamefont {Mizuno}},
  \bibinfo {author} {\bibfnamefont {M.}~\bibnamefont {Ochi}}, \bibinfo {author}
  {\bibfnamefont {H.}~\bibnamefont {Usui}},\ and\ \bibinfo {author}
  {\bibfnamefont {K.}~\bibnamefont {Kuroki}},\ }\bibfield  {title} {\bibinfo
  {title} {Theoretical study on the possibility of high ${T}_{c}$ $s\,\pm$-wave
  superconductivity in heavily hole-doped infinite layer nickelates},\ }\href
  {https://doi.org/10.1103/nl4m-n5dx} {\bibfield  {journal} {\bibinfo
  {journal} {Phys. Rev. B}\ }\textbf {\bibinfo {volume} {111}},\ \bibinfo
  {pages} {224511} (\bibinfo {year} {2025})}\BibitemShut {NoStop}%
\bibitem [{\citenamefont {Shinaoka}\ \emph {et~al.}(2015)\citenamefont
  {Shinaoka}, \citenamefont {Nomura}, \citenamefont {Biermann}, \citenamefont
  {Troyer},\ and\ \citenamefont {Werner}}]{H_Shinaoka_2015}%
  \BibitemOpen
  \bibfield  {author} {\bibinfo {author} {\bibfnamefont {H.}~\bibnamefont
  {Shinaoka}}, \bibinfo {author} {\bibfnamefont {Y.}~\bibnamefont {Nomura}},
  \bibinfo {author} {\bibfnamefont {S.}~\bibnamefont {Biermann}}, \bibinfo
  {author} {\bibfnamefont {M.}~\bibnamefont {Troyer}},\ and\ \bibinfo {author}
  {\bibfnamefont {P.}~\bibnamefont {Werner}},\ }\bibfield  {title} {\bibinfo
  {title} {Negative sign problem in continuous-time quantum monte carlo:
  Optimal choice of single-particle basis for impurity problems},\ }\href
  {https://doi.org/10.1103/PhysRevB.92.195126} {\bibfield  {journal} {\bibinfo
  {journal} {Phys. Rev. B}\ }\textbf {\bibinfo {volume} {92}},\ \bibinfo
  {pages} {195126} (\bibinfo {year} {2015})}\BibitemShut {NoStop}%
\bibitem [{\citenamefont {Poltavets}\ \emph {et~al.}(2006)\citenamefont
  {Poltavets}, \citenamefont {Lokshin}, \citenamefont {Dikmen}, \citenamefont
  {Croft}, \citenamefont {Egami},\ and\ \citenamefont
  {Greenblatt}}]{V_Poltavets_2006}%
  \BibitemOpen
  \bibfield  {author} {\bibinfo {author} {\bibfnamefont {V.~V.}\ \bibnamefont
  {Poltavets}}, \bibinfo {author} {\bibfnamefont {K.~A.}\ \bibnamefont
  {Lokshin}}, \bibinfo {author} {\bibfnamefont {S.}~\bibnamefont {Dikmen}},
  \bibinfo {author} {\bibfnamefont {M.}~\bibnamefont {Croft}}, \bibinfo
  {author} {\bibfnamefont {T.}~\bibnamefont {Egami}},\ and\ \bibinfo {author}
  {\bibfnamefont {M.}~\bibnamefont {Greenblatt}},\ }\bibfield  {title}
  {\bibinfo {title} {\ce{La3Ni2O6}: A new double ${T}'$-type nickelate with
  infinite \ce{Ni^{1+/2+}} layers},\ }\href {https://doi.org/10.1021/ja063031o}
  {\bibfield  {journal} {\bibinfo  {journal} {J. Am. Chem. Soc.}\ }\textbf
  {\bibinfo {volume} {128}},\ \bibinfo {pages} {9050} (\bibinfo {year}
  {2006})}\BibitemShut {NoStop}%
\bibitem [{\citenamefont {Poltavets}\ \emph {et~al.}(2009)\citenamefont
  {Poltavets}, \citenamefont {Greenblatt}, \citenamefont {Fecher},\ and\
  \citenamefont {Felser}}]{V_Poltavets_2009}%
  \BibitemOpen
  \bibfield  {author} {\bibinfo {author} {\bibfnamefont {V.~V.}\ \bibnamefont
  {Poltavets}}, \bibinfo {author} {\bibfnamefont {M.}~\bibnamefont
  {Greenblatt}}, \bibinfo {author} {\bibfnamefont {G.~H.}\ \bibnamefont
  {Fecher}},\ and\ \bibinfo {author} {\bibfnamefont {C.}~\bibnamefont
  {Felser}},\ }\bibfield  {title} {\bibinfo {title} {Electronic properties,
  band structure, and fermi surface instabilities of \ce{Ni^{1+/2+}} nickelate
  \ce{La3Ni2O6}, isoelectronic with superconducting cuprates},\ }\href
  {https://link.aps.org/doi/10.1103/PhysRevLett.102.046405} {\bibfield
  {journal} {\bibinfo  {journal} {Phys. Rev. Lett.}\ }\textbf {\bibinfo
  {volume} {102}},\ \bibinfo {pages} {046405} (\bibinfo {year}
  {2009})}\BibitemShut {NoStop}%
\bibitem [{\citenamefont {apRoberts Warren}\ \emph {et~al.}(2013)\citenamefont
  {apRoberts Warren}, \citenamefont {Crocker}, \citenamefont {Dioguardi},
  \citenamefont {Shirer}, \citenamefont {Poltavets}, \citenamefont
  {Greenblatt}, \citenamefont {Klavins},\ and\ \citenamefont
  {Curro}}]{N_Warren_2013}%
  \BibitemOpen
  \bibfield  {author} {\bibinfo {author} {\bibfnamefont {N.}~\bibnamefont
  {apRoberts Warren}}, \bibinfo {author} {\bibfnamefont {J.}~\bibnamefont
  {Crocker}}, \bibinfo {author} {\bibfnamefont {A.~P.}\ \bibnamefont
  {Dioguardi}}, \bibinfo {author} {\bibfnamefont {K.~R.}\ \bibnamefont
  {Shirer}}, \bibinfo {author} {\bibfnamefont {V.~V.}\ \bibnamefont
  {Poltavets}}, \bibinfo {author} {\bibfnamefont {M.}~\bibnamefont
  {Greenblatt}}, \bibinfo {author} {\bibfnamefont {P.}~\bibnamefont
  {Klavins}},\ and\ \bibinfo {author} {\bibfnamefont {N.~J.}\ \bibnamefont
  {Curro}},\ }\bibfield  {title} {\bibinfo {title} {Nmr evidence for spin
  fluctuations in the bilayer nickelate \ce{La3Ni2O6}},\ }\href
  {https://doi.org/10.1103/PhysRevB.88.075124} {\bibfield  {journal} {\bibinfo
  {journal} {Phys. Rev. B}\ }\textbf {\bibinfo {volume} {88}},\ \bibinfo
  {pages} {075124} (\bibinfo {year} {2013})}\BibitemShut {NoStop}%
\bibitem [{\citenamefont {Liu}\ \emph {et~al.}(2022)\citenamefont {Liu},
  \citenamefont {Sun}, \citenamefont {Huo}, \citenamefont {Ma}, \citenamefont
  {Ji}, \citenamefont {Yi}, \citenamefont {Li}, \citenamefont {Liu},
  \citenamefont {Yu}, \citenamefont {Zhang}, \citenamefont {Chen},
  \citenamefont {Liang}, \citenamefont {Dong}, \citenamefont {Guo},
  \citenamefont {Zhong}, \citenamefont {Shen}, \citenamefont {Li},\ and\
  \citenamefont {Wang}}]{Z_Liu_2022}%
  \BibitemOpen
  \bibfield  {author} {\bibinfo {author} {\bibfnamefont {Z.}~\bibnamefont
  {Liu}}, \bibinfo {author} {\bibfnamefont {H.}~\bibnamefont {Sun}}, \bibinfo
  {author} {\bibfnamefont {M.}~\bibnamefont {Huo}}, \bibinfo {author}
  {\bibfnamefont {X.}~\bibnamefont {Ma}}, \bibinfo {author} {\bibfnamefont
  {Y.}~\bibnamefont {Ji}}, \bibinfo {author} {\bibfnamefont {E.}~\bibnamefont
  {Yi}}, \bibinfo {author} {\bibfnamefont {L.}~\bibnamefont {Li}}, \bibinfo
  {author} {\bibfnamefont {H.}~\bibnamefont {Liu}}, \bibinfo {author}
  {\bibfnamefont {J.}~\bibnamefont {Yu}}, \bibinfo {author} {\bibfnamefont
  {Z.}~\bibnamefont {Zhang}}, \bibinfo {author} {\bibfnamefont
  {Z.}~\bibnamefont {Chen}}, \bibinfo {author} {\bibfnamefont {F.}~\bibnamefont
  {Liang}}, \bibinfo {author} {\bibfnamefont {H.}~\bibnamefont {Dong}},
  \bibinfo {author} {\bibfnamefont {H.}~\bibnamefont {Guo}}, \bibinfo {author}
  {\bibfnamefont {D.}~\bibnamefont {Zhong}}, \bibinfo {author} {\bibfnamefont
  {B.}~\bibnamefont {Shen}}, \bibinfo {author} {\bibfnamefont {S.}~\bibnamefont
  {Li}},\ and\ \bibinfo {author} {\bibfnamefont {M.}~\bibnamefont {Wang}},\
  }\bibfield  {title} {\bibinfo {title} {Evidence for charge and spin density
  waves in single crystals of \ce{La3Ni2O7} and \ce{La3Ni2O6}},\ }\href
  {https://doi.org/10.1007/s11433-022-1962-4} {\bibfield  {journal} {\bibinfo
  {journal} {Sci. China Phys. Mech. Astron.}\ }\textbf {\bibinfo {volume}
  {66}},\ \bibinfo {pages} {217411} (\bibinfo {year} {2022})}\BibitemShut
  {NoStop}%
\bibitem [{\citenamefont {Botana}\ \emph {et~al.}(2016)\citenamefont {Botana},
  \citenamefont {Pardo}, \citenamefont {Pickett},\ and\ \citenamefont
  {Norman}}]{A_Botana_2016}%
  \BibitemOpen
  \bibfield  {author} {\bibinfo {author} {\bibfnamefont {A.~S.}\ \bibnamefont
  {Botana}}, \bibinfo {author} {\bibfnamefont {V.}~\bibnamefont {Pardo}},
  \bibinfo {author} {\bibfnamefont {W.~E.}\ \bibnamefont {Pickett}},\ and\
  \bibinfo {author} {\bibfnamefont {M.~R.}\ \bibnamefont {Norman}},\ }\bibfield
   {title} {\bibinfo {title} {Charge ordering in \ce{Ni^{1+}/Ni^{2+}}
  nickelates: \ce{La4Ni3O8} and \ce{La3Ni2O6}},\ }\href
  {https://doi.org/10.1103/PhysRevB.94.081105} {\bibfield  {journal} {\bibinfo
  {journal} {Phys. Rev. B}\ }\textbf {\bibinfo {volume} {94}},\ \bibinfo
  {pages} {081105} (\bibinfo {year} {2016})}\BibitemShut {NoStop}%
\bibitem [{\citenamefont {Worm}\ \emph {et~al.}(2022)\citenamefont {Worm},
  \citenamefont {Si}, \citenamefont {Kitatani}, \citenamefont {Arita},
  \citenamefont {Tomczak},\ and\ \citenamefont {Held}}]{P_Worm_2022}%
  \BibitemOpen
  \bibfield  {author} {\bibinfo {author} {\bibfnamefont {P.}~\bibnamefont
  {Worm}}, \bibinfo {author} {\bibfnamefont {L.}~\bibnamefont {Si}}, \bibinfo
  {author} {\bibfnamefont {M.}~\bibnamefont {Kitatani}}, \bibinfo {author}
  {\bibfnamefont {R.}~\bibnamefont {Arita}}, \bibinfo {author} {\bibfnamefont
  {J.~M.}\ \bibnamefont {Tomczak}},\ and\ \bibinfo {author} {\bibfnamefont
  {K.}~\bibnamefont {Held}},\ }\bibfield  {title} {\bibinfo {title}
  {Correlations tune the electronic structure of pentalayer nickelates into the
  superconducting regime},\ }\href
  {https://doi.org/10.1103/PhysRevMaterials.6.L091801} {\bibfield  {journal}
  {\bibinfo  {journal} {Phys. Rev. Mater.}\ }\textbf {\bibinfo {volume} {6}},\
  \bibinfo {pages} {L091801} (\bibinfo {year} {2022})}\BibitemShut {NoStop}%
\bibitem [{\citenamefont {Zhang}\ \emph {et~al.}(2024)\citenamefont {Zhang},
  \citenamefont {Lin}, \citenamefont {Moreo}, \citenamefont {Maier},\ and\
  \citenamefont {Dagotto}}]{Y_Zhang_2024}%
  \BibitemOpen
  \bibfield  {author} {\bibinfo {author} {\bibfnamefont {Y.}~\bibnamefont
  {Zhang}}, \bibinfo {author} {\bibfnamefont {L.-F.}\ \bibnamefont {Lin}},
  \bibinfo {author} {\bibfnamefont {A.}~\bibnamefont {Moreo}}, \bibinfo
  {author} {\bibfnamefont {T.~A.}\ \bibnamefont {Maier}},\ and\ \bibinfo
  {author} {\bibfnamefont {E.}~\bibnamefont {Dagotto}},\ }\bibfield  {title}
  {\bibinfo {title} {Electronic structure, magnetic correlations, and
  superconducting pairing in the reduced ruddlesden-popper bilayer
  \ce{La3Ni2O6} under pressure: Different role of
  ${d}_{3{z}^{2}\ensuremath{-}{r}^{2}}$ orbital compared with \ce{La3Ni2O7}},\
  }\href {https://doi.org/10.1103/PhysRevB.109.045151} {\bibfield  {journal}
  {\bibinfo  {journal} {Phys. Rev. B}\ }\textbf {\bibinfo {volume} {109}},\
  \bibinfo {pages} {045151} (\bibinfo {year} {2024})}\BibitemShut {NoStop}%
\bibitem [{\citenamefont {M^^c3^^bcller-Buschbaum}(1977)}]{H_Muller_1977}%
  \BibitemOpen
  \bibfield  {author} {\bibinfo {author} {\bibfnamefont {H.}~\bibnamefont
  {M^^c3^^bcller-Buschbaum}},\ }\bibfield  {title} {\bibinfo {title}
  {Oxometallates with planar coordination},\ }\href
  {https://doi.org/https://doi.org/10.1002/anie.197706741} {\bibfield
  {journal} {\bibinfo  {journal} {Angew. Chem. Int. Ed. Engl.}\ }\textbf
  {\bibinfo {volume} {16}},\ \bibinfo {pages} {674} (\bibinfo {year}
  {1977})}\BibitemShut {NoStop}%
\bibitem [{\citenamefont {Lechermann}\ \emph {et~al.}(2025)\citenamefont
  {Lechermann}, \citenamefont {B\"otzel},\ and\ \citenamefont
  {Eremin}}]{F_Lechermann_2024}%
  \BibitemOpen
  \bibfield  {author} {\bibinfo {author} {\bibfnamefont {F.}~\bibnamefont
  {Lechermann}}, \bibinfo {author} {\bibfnamefont {S.}~\bibnamefont
  {B\"otzel}},\ and\ \bibinfo {author} {\bibfnamefont {I.~M.}\ \bibnamefont
  {Eremin}},\ }\bibfield  {title} {\bibinfo {title} {Interplay of
  orbital-selective mott criticality and flat-band physics in \ce{La3Ni2O6}},\
  }\href {https://doi.org/10.1103/3r7g-89r8} {\bibfield  {journal} {\bibinfo
  {journal} {Phys. Rev. B}\ }\textbf {\bibinfo {volume} {112}},\ \bibinfo
  {pages} {245125} (\bibinfo {year} {2025})}\BibitemShut {NoStop}%
\bibitem [{\citenamefont {Perdew}\ \emph {et~al.}(2008)\citenamefont {Perdew},
  \citenamefont {Ruzsinszky}, \citenamefont {Csonka}, \citenamefont {Vydrov},
  \citenamefont {Scuseria}, \citenamefont {Constantin}, \citenamefont {Zhou},\
  and\ \citenamefont {Burke}}]{PBEsol_1}%
  \BibitemOpen
  \bibfield  {author} {\bibinfo {author} {\bibfnamefont {J.~P.}\ \bibnamefont
  {Perdew}}, \bibinfo {author} {\bibfnamefont {A.}~\bibnamefont {Ruzsinszky}},
  \bibinfo {author} {\bibfnamefont {G.~I.}\ \bibnamefont {Csonka}}, \bibinfo
  {author} {\bibfnamefont {O.~A.}\ \bibnamefont {Vydrov}}, \bibinfo {author}
  {\bibfnamefont {G.~E.}\ \bibnamefont {Scuseria}}, \bibinfo {author}
  {\bibfnamefont {L.~A.}\ \bibnamefont {Constantin}}, \bibinfo {author}
  {\bibfnamefont {X.}~\bibnamefont {Zhou}},\ and\ \bibinfo {author}
  {\bibfnamefont {K.}~\bibnamefont {Burke}},\ }\bibfield  {title} {\bibinfo
  {title} {Restoring the density-gradient expansion for exchange in solids and
  surfaces},\ }\href {https://doi.org/10.1103/PhysRevLett.100.136406}
  {\bibfield  {journal} {\bibinfo  {journal} {Phys. Rev. Lett.}\ }\textbf
  {\bibinfo {volume} {100}},\ \bibinfo {pages} {136406} (\bibinfo {year}
  {2008})}\BibitemShut {NoStop}%
\bibitem [{\citenamefont {Perdew}\ \emph {et~al.}(2009)\citenamefont {Perdew},
  \citenamefont {Ruzsinszky}, \citenamefont {Csonka}, \citenamefont {Vydrov},
  \citenamefont {Scuseria}, \citenamefont {Constantin}, \citenamefont {Zhou},\
  and\ \citenamefont {Burke}}]{PBEsol_2}%
  \BibitemOpen
  \bibfield  {author} {\bibinfo {author} {\bibfnamefont {J.~P.}\ \bibnamefont
  {Perdew}}, \bibinfo {author} {\bibfnamefont {A.}~\bibnamefont {Ruzsinszky}},
  \bibinfo {author} {\bibfnamefont {G.~I.}\ \bibnamefont {Csonka}}, \bibinfo
  {author} {\bibfnamefont {O.~A.}\ \bibnamefont {Vydrov}}, \bibinfo {author}
  {\bibfnamefont {G.~E.}\ \bibnamefont {Scuseria}}, \bibinfo {author}
  {\bibfnamefont {L.~A.}\ \bibnamefont {Constantin}}, \bibinfo {author}
  {\bibfnamefont {X.}~\bibnamefont {Zhou}},\ and\ \bibinfo {author}
  {\bibfnamefont {K.}~\bibnamefont {Burke}},\ }\bibfield  {title} {\bibinfo
  {title} {Erratum: Restoring the density-gradient expansion for exchange in
  solids and surfaces},\ }\href
  {https://doi.org/10.1103/PhysRevLett.102.039902} {\bibfield  {journal}
  {\bibinfo  {journal} {Phys. Rev. Lett.}\ }\textbf {\bibinfo {volume} {102}},\
  \bibinfo {pages} {039902} (\bibinfo {year} {2009})}\BibitemShut {NoStop}%
\bibitem [{\citenamefont {Kresse}\ and\ \citenamefont {Hafner}(1993)}]{vasp1}%
  \BibitemOpen
  \bibfield  {author} {\bibinfo {author} {\bibfnamefont {G.}~\bibnamefont
  {Kresse}}\ and\ \bibinfo {author} {\bibfnamefont {J.}~\bibnamefont
  {Hafner}},\ }\bibfield  {title} {\bibinfo {title} {Ab initio molecular
  dynamics for liquid metals},\ }\href
  {https://doi.org/10.1103/PhysRevB.47.558} {\bibfield  {journal} {\bibinfo
  {journal} {Phys. Rev. B}\ }\textbf {\bibinfo {volume} {47}},\ \bibinfo
  {pages} {558} (\bibinfo {year} {1993})}\BibitemShut {NoStop}%
\bibitem [{\citenamefont {Kresse}\ and\ \citenamefont {Hafner}(1994)}]{vasp2}%
  \BibitemOpen
  \bibfield  {author} {\bibinfo {author} {\bibfnamefont {G.}~\bibnamefont
  {Kresse}}\ and\ \bibinfo {author} {\bibfnamefont {J.}~\bibnamefont
  {Hafner}},\ }\bibfield  {title} {\bibinfo {title} {Ab initio
  molecular-dynamics simulation of the liquid-metal--amorphous-semiconductor
  transition in germanium},\ }\href {https://doi.org/10.1103/PhysRevB.49.14251}
  {\bibfield  {journal} {\bibinfo  {journal} {Phys. Rev. B}\ }\textbf {\bibinfo
  {volume} {49}},\ \bibinfo {pages} {14251} (\bibinfo {year}
  {1994})}\BibitemShut {NoStop}%
\bibitem [{\citenamefont {Kresse}\ and\ \citenamefont
  {Furthm{\"u}ller}(1996)}]{vasp3}%
  \BibitemOpen
  \bibfield  {author} {\bibinfo {author} {\bibfnamefont {G.}~\bibnamefont
  {Kresse}}\ and\ \bibinfo {author} {\bibfnamefont {J.}~\bibnamefont
  {Furthm{\"u}ller}},\ }\bibfield  {title} {\bibinfo {title} {Efficiency of
  ab-initio total energy calculations for metals and semiconductors using a
  plane-wave basis set},\ }\href
  {https://doi.org/https://doi.org/10.1016/0927-0256(96)00008-0} {\bibfield
  {journal} {\bibinfo  {journal} {Comput. Mater. Sci.}\ }\textbf {\bibinfo
  {volume} {6}},\ \bibinfo {pages} {15} (\bibinfo {year} {1996})}\BibitemShut
  {NoStop}%
\bibitem [{\citenamefont {Kresse}\ and\ \citenamefont
  {Furthm\"uller}(1996)}]{vasp}%
  \BibitemOpen
  \bibfield  {author} {\bibinfo {author} {\bibfnamefont {G.}~\bibnamefont
  {Kresse}}\ and\ \bibinfo {author} {\bibfnamefont {J.}~\bibnamefont
  {Furthm\"uller}},\ }\bibfield  {title} {\bibinfo {title} {Efficient iterative
  schemes for ab initio total-energy calculations using a plane-wave basis
  set},\ }\href {https://doi.org/10.1103/PhysRevB.54.11169} {\bibfield
  {journal} {\bibinfo  {journal} {Phys. Rev. B}\ }\textbf {\bibinfo {volume}
  {54}},\ \bibinfo {pages} {11169} (\bibinfo {year} {1996})}\BibitemShut
  {NoStop}%
\bibitem [{\citenamefont {Kresse}\ and\ \citenamefont {Joubert}(1999)}]{vasp4}%
  \BibitemOpen
  \bibfield  {author} {\bibinfo {author} {\bibfnamefont {G.}~\bibnamefont
  {Kresse}}\ and\ \bibinfo {author} {\bibfnamefont {D.}~\bibnamefont
  {Joubert}},\ }\bibfield  {title} {\bibinfo {title} {From ultrasoft
  pseudopotentials to the projector augmented-wave method},\ }\href
  {https://doi.org/10.1103/PhysRevB.59.1758} {\bibfield  {journal} {\bibinfo
  {journal} {Phys. Rev. B}\ }\textbf {\bibinfo {volume} {59}},\ \bibinfo
  {pages} {1758} (\bibinfo {year} {1999})}\BibitemShut {NoStop}%
\bibitem [{\citenamefont {Kotani}(2014)}]{Kotani2014PMT-QSGW}%
  \BibitemOpen
  \bibfield  {author} {\bibinfo {author} {\bibfnamefont {T.}~\bibnamefont
  {Kotani}},\ }\bibfield  {title} {\bibinfo {title} {Quasiparticle
  self-consistent {G}{W} method based on the augmented plane-wave and
  muffin-tin orbital method},\ }\href {https://doi.org/10.7566/JPSJ.83.094711}
  {\bibfield  {journal} {\bibinfo  {journal} {J. Phys. Soc. Jpn.}\ }\textbf
  {\bibinfo {volume} {83}},\ \bibinfo {pages} {094711} (\bibinfo {year}
  {2014})}\BibitemShut {NoStop}%
\bibitem [{\citenamefont {Kotani}\ and\ \citenamefont {van
  Schilfgaarde}(2010)}]{Kotani2010PMT}%
  \BibitemOpen
  \bibfield  {author} {\bibinfo {author} {\bibfnamefont {T.}~\bibnamefont
  {Kotani}}\ and\ \bibinfo {author} {\bibfnamefont {M.}~\bibnamefont {van
  Schilfgaarde}},\ }\bibfield  {title} {\bibinfo {title} {Fusion of the lapw
  and the lmto methods: The augmented plane wave plus muffin-tin orbital (pmt)
  method},\ }\href {https://doi.org/10.1103/PhysRevB.81.125117} {\bibfield
  {journal} {\bibinfo  {journal} {Phys. Rev. B}\ }\textbf {\bibinfo {volume}
  {81}},\ \bibinfo {pages} {125117} (\bibinfo {year} {2010})}\BibitemShut
  {NoStop}%
\bibitem [{\citenamefont {Kotani}\ \emph {et~al.}(2007)\citenamefont {Kotani},
  \citenamefont {van Schilfgaarde},\ and\ \citenamefont
  {Faleev}}]{Kotani2007QSGW}%
  \BibitemOpen
  \bibfield  {author} {\bibinfo {author} {\bibfnamefont {T.}~\bibnamefont
  {Kotani}}, \bibinfo {author} {\bibfnamefont {M.}~\bibnamefont {van
  Schilfgaarde}},\ and\ \bibinfo {author} {\bibfnamefont {S.~V.}\ \bibnamefont
  {Faleev}},\ }\bibfield  {title} {\bibinfo {title} {Quasiparticle
  self-consistent {G}{W} method: A basis for the independent-particle
  approximation},\ }\href {https://doi.org/10.1103/PhysRevB.76.165106}
  {\bibfield  {journal} {\bibinfo  {journal} {Phys. Rev. B}\ }\textbf {\bibinfo
  {volume} {76}},\ \bibinfo {pages} {165106} (\bibinfo {year}
  {2007})}\BibitemShut {NoStop}%
\bibitem [{eca()}]{ecalj}%
  \BibitemOpen
  \href@noop {} {\bibinfo {title} {{ecalj}: A first-principles
  electronic-structure suite based on the {PMT} method}},\ \bibinfo
  {howpublished} {\url{https://github.com/tkotani/ecalj}},\ \bibinfo {note}
  {the one-body part is developed based on the {LMsuit} package
  (\url{http://www.lmsuite.org/})}\BibitemShut {NoStop}%
\bibitem [{\citenamefont {Deguchi}\ \emph {et~al.}(2016)\citenamefont
  {Deguchi}, \citenamefont {Sato}, \citenamefont {Kino},\ and\ \citenamefont
  {Kotani}}]{Deguchi2016QSGW80}%
  \BibitemOpen
  \bibfield  {author} {\bibinfo {author} {\bibfnamefont {D.}~\bibnamefont
  {Deguchi}}, \bibinfo {author} {\bibfnamefont {K.}~\bibnamefont {Sato}},
  \bibinfo {author} {\bibfnamefont {H.}~\bibnamefont {Kino}},\ and\ \bibinfo
  {author} {\bibfnamefont {T.}~\bibnamefont {Kotani}},\ }\bibfield  {title}
  {\bibinfo {title} {Accurate energy bands calculated by the hybrid
  quasiparticle self-consistent {G}{W} method implemented in the ecalj
  package},\ }\href {https://doi.org/10.7567/JJAP.55.051201} {\bibfield
  {journal} {\bibinfo  {journal} {Jpn. J. Appl. Phys.}\ }\textbf {\bibinfo
  {volume} {55}},\ \bibinfo {pages} {051201} (\bibinfo {year}
  {2016})}\BibitemShut {NoStop}%
\bibitem [{\citenamefont {Marzari}\ and\ \citenamefont
  {Vanderbilt}(1997)}]{wannier_1}%
  \BibitemOpen
  \bibfield  {author} {\bibinfo {author} {\bibfnamefont {N.}~\bibnamefont
  {Marzari}}\ and\ \bibinfo {author} {\bibfnamefont {D.}~\bibnamefont
  {Vanderbilt}},\ }\bibfield  {title} {\bibinfo {title} {Maximally localized
  generalized wannier functions for composite energy bands},\ }\href
  {https://doi.org/10.1103/PhysRevB.56.12847} {\bibfield  {journal} {\bibinfo
  {journal} {Phys. Rev. B}\ }\textbf {\bibinfo {volume} {56}},\ \bibinfo
  {pages} {12847} (\bibinfo {year} {1997})}\BibitemShut {NoStop}%
\bibitem [{\citenamefont {Souza}\ \emph {et~al.}(2001)\citenamefont {Souza},
  \citenamefont {Marzari},\ and\ \citenamefont {Vanderbilt}}]{wannier_2}%
  \BibitemOpen
  \bibfield  {author} {\bibinfo {author} {\bibfnamefont {I.}~\bibnamefont
  {Souza}}, \bibinfo {author} {\bibfnamefont {N.}~\bibnamefont {Marzari}},\
  and\ \bibinfo {author} {\bibfnamefont {D.}~\bibnamefont {Vanderbilt}},\
  }\bibfield  {title} {\bibinfo {title} {Maximally localized wannier functions
  for entangled energy bands},\ }\href
  {https://doi.org/10.1103/PhysRevB.65.035109} {\bibfield  {journal} {\bibinfo
  {journal} {Phys. Rev. B}\ }\textbf {\bibinfo {volume} {65}},\ \bibinfo
  {pages} {035109} (\bibinfo {year} {2001})}\BibitemShut {NoStop}%
\bibitem [{\citenamefont {Mostofi}\ \emph {et~al.}(2008)\citenamefont
  {Mostofi}, \citenamefont {Yates}, \citenamefont {Lee}, \citenamefont {Souza},
  \citenamefont {Vanderbilt},\ and\ \citenamefont {Marzari}}]{wannier90}%
  \BibitemOpen
  \bibfield  {author} {\bibinfo {author} {\bibfnamefont {A.~A.}\ \bibnamefont
  {Mostofi}}, \bibinfo {author} {\bibfnamefont {J.~R.}\ \bibnamefont {Yates}},
  \bibinfo {author} {\bibfnamefont {Y.-S.}\ \bibnamefont {Lee}}, \bibinfo
  {author} {\bibfnamefont {I.}~\bibnamefont {Souza}}, \bibinfo {author}
  {\bibfnamefont {D.}~\bibnamefont {Vanderbilt}},\ and\ \bibinfo {author}
  {\bibfnamefont {N.}~\bibnamefont {Marzari}},\ }\bibfield  {title} {\bibinfo
  {title} {wannier90: A tool for obtaining maximally-localised wannier
  functions},\ }\href
  {https://doi.org/https://doi.org/10.1016/j.cpc.2007.11.016} {\bibfield
  {journal} {\bibinfo  {journal} {Comput. Phys. Commun.}\ }\textbf {\bibinfo
  {volume} {178}},\ \bibinfo {pages} {685} (\bibinfo {year}
  {2008})}\BibitemShut {NoStop}%
\bibitem [{\citenamefont {Bickers}\ \emph {et~al.}(1989)\citenamefont
  {Bickers}, \citenamefont {Scalapino},\ and\ \citenamefont {White}}]{FLEX_1}%
  \BibitemOpen
  \bibfield  {author} {\bibinfo {author} {\bibfnamefont {N.~E.}\ \bibnamefont
  {Bickers}}, \bibinfo {author} {\bibfnamefont {D.~J.}\ \bibnamefont
  {Scalapino}},\ and\ \bibinfo {author} {\bibfnamefont {S.~R.}\ \bibnamefont
  {White}},\ }\bibfield  {title} {\bibinfo {title} {Conserving approximations
  for strongly correlated electron systems: Bethe-salpeter equation and
  dynamics for the two-dimensional hubbard model},\ }\href
  {https://doi.org/10.1103/PhysRevLett.62.961} {\bibfield  {journal} {\bibinfo
  {journal} {Phys. Rev. Lett.}\ }\textbf {\bibinfo {volume} {62}},\ \bibinfo
  {pages} {961} (\bibinfo {year} {1989})}\BibitemShut {NoStop}%
\bibitem [{\citenamefont {Bickers}\ and\ \citenamefont {White}(1991)}]{FLEX_2}%
  \BibitemOpen
  \bibfield  {author} {\bibinfo {author} {\bibfnamefont {N.~E.}\ \bibnamefont
  {Bickers}}\ and\ \bibinfo {author} {\bibfnamefont {S.~R.}\ \bibnamefont
  {White}},\ }\bibfield  {title} {\bibinfo {title} {Conserving approximations
  for strongly fluctuating electron systems. ii. numerical results and parquet
  extension},\ }\href {https://doi.org/10.1103/PhysRevB.43.8044} {\bibfield
  {journal} {\bibinfo  {journal} {Phys. Rev. B}\ }\textbf {\bibinfo {volume}
  {43}},\ \bibinfo {pages} {8044} (\bibinfo {year} {1991})}\BibitemShut
  {NoStop}%
\bibitem [{\citenamefont {Togo}\ \emph {et~al.}(2023)\citenamefont {Togo},
  \citenamefont {Chaput}, \citenamefont {Tadano},\ and\ \citenamefont
  {Tanaka}}]{phonopy_1}%
  \BibitemOpen
  \bibfield  {author} {\bibinfo {author} {\bibfnamefont {A.}~\bibnamefont
  {Togo}}, \bibinfo {author} {\bibfnamefont {L.}~\bibnamefont {Chaput}},
  \bibinfo {author} {\bibfnamefont {T.}~\bibnamefont {Tadano}},\ and\ \bibinfo
  {author} {\bibfnamefont {I.}~\bibnamefont {Tanaka}},\ }\bibfield  {title}
  {\bibinfo {title} {Implementation strategies in phonopy and phono3py},\
  }\href {https://doi.org/10.1088/1361-648X/acd831} {\bibfield  {journal}
  {\bibinfo  {journal} {J. Phys.: Condens. Matter}\ }\textbf {\bibinfo {volume}
  {35}},\ \bibinfo {pages} {353001} (\bibinfo {year} {2023})}\BibitemShut
  {NoStop}%
\bibitem [{\citenamefont {Togo}(2023)}]{phonopy_2}%
  \BibitemOpen
  \bibfield  {author} {\bibinfo {author} {\bibfnamefont {A.}~\bibnamefont
  {Togo}},\ }\bibfield  {title} {\bibinfo {title} {First-principles phonon
  calculations with phonopy and phono3py},\ }\href
  {https://doi.org/10.7566/JPSJ.92.012001} {\bibfield  {journal} {\bibinfo
  {journal} {J. Phys. Soc. Jpn.}\ }\textbf {\bibinfo {volume} {92}},\ \bibinfo
  {pages} {012001} (\bibinfo {year} {2023})}\BibitemShut {NoStop}%
\bibitem [{\citenamefont {Jang}\ \emph {et~al.}(2015)\citenamefont {Jang},
  \citenamefont {Kotani}, \citenamefont {Kino}, \citenamefont {Kuroki},\ and\
  \citenamefont {Han}}]{Q_Jang_2015}%
  \BibitemOpen
  \bibfield  {author} {\bibinfo {author} {\bibfnamefont {S.~W.}\ \bibnamefont
  {Jang}}, \bibinfo {author} {\bibfnamefont {T.}~\bibnamefont {Kotani}},
  \bibinfo {author} {\bibfnamefont {H.}~\bibnamefont {Kino}}, \bibinfo {author}
  {\bibfnamefont {K.}~\bibnamefont {Kuroki}},\ and\ \bibinfo {author}
  {\bibfnamefont {M.~J.}\ \bibnamefont {Han}},\ }\bibfield  {title} {\bibinfo
  {title} {Quasiparticle self-consistent gw study of cuprates: electronic
  structure, model parameters and the two-band theory for ${T}_c$},\ }\href
  {https://doi.org/10.1038/srep12050} {\bibfield  {journal} {\bibinfo
  {journal} {Sci. Rep.}\ }\textbf {\bibinfo {volume} {5}},\ \bibinfo {pages}
  {12050} (\bibinfo {year} {2015})}\BibitemShut {NoStop}%
\bibitem [{\citenamefont {Hoshi}\ \emph {et~al.}(2026)\citenamefont {Hoshi},
  \citenamefont {Kuroki},\ and\ \citenamefont {Sakakibrara}}]{H_Yuto_2026}%
  \BibitemOpen
  \bibfield  {author} {\bibinfo {author} {\bibfnamefont {Y.}~\bibnamefont
  {Hoshi}}, \bibinfo {author} {\bibfnamefont {K.}~\bibnamefont {Kuroki}},\ and\
  \bibinfo {author} {\bibfnamefont {H.}~\bibnamefont {Sakakibrara}}} (\bibinfo
  {year} {2026}),\ \bibinfo {note} {in preparation}\BibitemShut {NoStop}%
\bibitem [{\citenamefont {Sakakibara}\ \emph {et~al.}(2024)\citenamefont
  {Sakakibara}, \citenamefont {Kitamine}, \citenamefont {Ochi},\ and\
  \citenamefont {Kuroki}}]{H_Sakakibara_2024}%
  \BibitemOpen
  \bibfield  {author} {\bibinfo {author} {\bibfnamefont {H.}~\bibnamefont
  {Sakakibara}}, \bibinfo {author} {\bibfnamefont {N.}~\bibnamefont
  {Kitamine}}, \bibinfo {author} {\bibfnamefont {M.}~\bibnamefont {Ochi}},\
  and\ \bibinfo {author} {\bibfnamefont {K.}~\bibnamefont {Kuroki}},\
  }\bibfield  {title} {\bibinfo {title} {Possible high ${T}_{c}$
  superconductivity in \ce{La3Ni2O7} under high pressure through manifestation
  of a nearly half-filled bilayer hubbard model},\ }\href
  {https://doi.org/10.1103/PhysRevLett.132.106002} {\bibfield  {journal}
  {\bibinfo  {journal} {Phys. Rev. Lett.}\ }\textbf {\bibinfo {volume} {132}},\
  \bibinfo {pages} {106002} (\bibinfo {year} {2024})}\BibitemShut {NoStop}%
\bibitem [{\citenamefont {Shannon}(1976)}]{R_Shannon_1976}%
  \BibitemOpen
  \bibfield  {author} {\bibinfo {author} {\bibfnamefont {R.~D.}\ \bibnamefont
  {Shannon}},\ }\bibfield  {title} {\bibinfo {title} {Revised effective ionic
  radii and systematic studies of interatomic distances in halides and
  chalcogenides},\ }\href
  {https://doi.org/https://doi.org/10.1107/S0567739476001551} {\bibfield
  {journal} {\bibinfo  {journal} {Acta Cryst. A}\ }\textbf {\bibinfo {volume}
  {32}},\ \bibinfo {pages} {751} (\bibinfo {year} {1976})}\BibitemShut
  {NoStop}%
\end{thebibliography}%

\end{document}